\documentclass[a4paper,12pt]{article}
\usepackage{indentfirst}
\usepackage{slashed}
\usepackage{amsfonts}
\usepackage{multirow}
\usepackage{mathrsfs}
\usepackage{graphicx}
\usepackage{amsmath}
\usepackage{amssymb}
\usepackage{bm}
\usepackage{bbm}
\usepackage{color}
\usepackage[numbers, square, comma, sort&compress]{natbib}
\usepackage{hyperref}
\usepackage{geometry}
\usepackage[utf8]{inputenc}
\usepackage{subfig} 
\usepackage{float}
\usepackage{comment}

\newcommand{\nn}{\nonumber}
\newcommand{\beq}{\begin{equation}}
\newcommand{\eeq}{\end{equation}}
\newcommand{\bqa}{\begin{eqnarray}}
\newcommand{\eqa}{\end{eqnarray}}

\def\ksl{k\!\!\!\slash}
\def\lsl{l\!\!\!\slash}
\def\nsl{n\!\!\!\slash}
\def\nbsl{\bar{n}\!\!\!\slash}

\allowdisplaybreaks[4]

\graphicspath{ {./diagram/}, {./diagram2/} }

\begin{document}
\begin{titlepage}
\vskip 25mm
\begin{center}
\Large\bf{Region analysis of  $H\to \gamma \gamma $ via a bottom quark loop }
\end{center}
\vskip 8mm

\begin{center}
{\bf Jun-Yao Hou$^a$, Jian Wang$^{a,b}$, Da-Jiang Zhang$^a$}\\
\vspace{10mm}
\textit{${}^a$School of Physics, Shandong University, Jinan, Shandong 250100, China}\\
\textit{${}^b$Center for High Energy Physics, Peking University, Beijing 100871, China}\\
\vspace{3mm}
\textit{Email:} j.wang@sdu.edu.cn
\vspace{5mm}
\end{center}

\begin{abstract}
The $H\to \gamma\gamma$ decay is an ideal process to study the structure of next-to-leading power logarithms induced by quarks due to its simple initial and final states.
We perform a region analysis of this process up to two-loop level to inspect the origins of the logarithms.
To deal with the endpoint singularities that are prevalent for the next-to-leading power logarithms, we have adopted two different kinds of regulators to exhibit the advantages and disadvantages of each regulator. 
In the analytic regulator we have chosen, the power of the propagator is changed by $\eta$.
And the endpoint singularities are regulated in the form of $1/\eta$.
These poles cancel between the collinear and anti-collinear sectors since there is no soft mode in such a regulator.
In the $\Delta$ regulator, 
the soft sector is important.
The leading and next-to-leading logarithms can be inferred from only this sector.
Moreover, the symmetry between the collinear and anti-collinear sectors is preserved.
After imposing a cut on the bottom quark transverse momentum, the leading order result is finite in each sector.
We also discuss the next-to-next-to-leading power contributions
and find that the potential factorization formulae involve two-dimensional endpoint singularities.
Our region analysis could help to develop sophisticated factorization and resummation schemes beyond leading power.

\end{abstract}

\vspace{10mm}

\end{titlepage}

\section{Introduction}

Precise predictions for the scattering processes at colliders play an important role in stringent tests of the standard model and search for new physics \cite{Heinrich:2020ybq,Boughezal:2022cbl}.
The relevant virtual and real corrections need to be computed to higher orders in perturbation theories, either analytically or numerically.
Along this line, impressive progresses have been made in the past decade.
An indispensable  ingredient in next-to-next-to-leading order (or even higher-order) corrections is the analytical calculation of  infrared divergences in the real corrections. 
It is convenient to construct this part by taking advantage of the knowledge of an effective field theory such as the soft-collinear effective theory \cite{Bauer:2001yt,Bauer:2002nz,Beneke:2002ph,Beneke:2002ni}.
To ensure the application of this effective theory, a kinematic cut, e.g., the transverse momentum $q_T$ or the $N$-jettiness variable \cite{Stewart:2010tn}, is usually imposed on the final states \cite{Catani:2007vq,Boughezal:2015dva,Gaunt:2015pea}.
The cross section above the cut is guaranteed to be free of infrared divergences and thus can be calculated numerically using Monte Carlo integration methods.
In practice, the Monte Carlo integration is  cumbersome and time-consuming.
It often happens that a program needs several months to accomplish a computation of the cross section under a specific setup.
The main reason is that the cut is usually chosen at such a small value that the cross section varies dramatically near the cut.
Therefore, it is beneficial to have a better understanding of the cross section below the cut \cite{Gehrmann:2012ze,Gehrmann:2014yya,Luo:2019hmp,Luo:2019bmw,Luo:2019szz,Ebert:2020yqt,Luo:2020epw,Becher:2006qw,Becher:2010pd,Gaunt:2014xga,Gaunt:2014cfa,Boughezal:2015eha,Li:2016tvb,Campbell:2017hsw,Li:2018tsq,Jin:2019dho,Bell:2023yso,Agarwal:2024gws,Bruser:2018rad,Banerjee:2018ozf,Ebert:2020unb,Baranowski:2022vcn,Baranowski:2022khd,Chen:2022yre,Baranowski:2024vxg,Baranowski:2024ysi,Buonocore:2023rdw,Fu:2024fgj}, especially the power corrections \cite{Moult:2016fqy,Boughezal:2016zws,Moult:2017jsg,Balitsky:2017gis,Ebert:2018gsn,Boughezal:2019ggi,Ebert:2020dfc,Oleari:2020wvt,Inglis-Whalen:2021bea,Ferrera:2023vsw,Vita:2024ypr}, so that one can choose a larger value of the cut.

Besides the phenomenological application, 
the power corrections for the cross sections are of great theoretical interest in its own right \cite{Beneke:2019kgv,Moult:2018jjd,Beneke:2018gvs,Moult:2019uhz,Beneke:2019mua,Beneke:2019oqx,Beneke:2020ibj,Beneke:2022obx,Bonocore:2016awd,DelDuca:2017twk,vanBeekveld:2019prq,Laenen:2020nrt,AH:2020iki,Pal:2023vec,Pal:2024eyr,Liu:2017vkm,Liu:2018czl,Liu:2019oav,Wang:2019mym,Liu:2020tzd,Liu:2020wbn,Anastasiou:2020vkr,Liu:2021chn,Liu:2022ajh,Bell:2022ott}.
One of the unique features is the appearance of the endpoint singularities in the factorization formula of the relevant cross section.
At leading power (LP), the hard scattering process is induced 
by an operator with only a single collinear building block, i.e., a gauge invariant collinear field, in each collinear direction \cite{Beneke:2017ztn,Beneke:2018rbh}.
As such, the short-distance hard function depends only on the total momentum of the collinear building block.
The long-distance soft and collinear functions rely on the low scale fluctuations that can be completely factorized from the hard function.
At next-to-leading power (NLP), however, the factorization scheme  becomes intricate.
Firstly, the hard process can be triggered by an operator with two collinear building blocks in a direction \cite{Beneke:2017ztn,Beneke:2018rbh}.
The hard function depends on each individual momentum of the building block in terms of its large light-cone component fraction $z$.
And due to the very multiple external states, the hard function often contains $1/z$ and $1/(1-z)$ singular terms, which arise from the presumed highly off-shell propagators.
Because any physical observable should not be sensitive to such a collinear structure, it is mandatory to integrate the $z$ variable in both the hard and collinear functions over all the allowed region, usually from $0$ to $1$, which causes endpoint singularities.
Although they can be regulated in the $D$-dimensional factorization formula, they prevent the resummation of large logarithms in the conventional way, i.e.,
performing renormalization of the various functions before taking the integration of $z$.
Secondly, the soft quark begins to make a contribution.
In contrast to the leading power soft gluon effects,
the contribution of the soft quark cannot be simply organized into a soft Wilson line.
Actually, it leads to the occurrence of a radiative jet function, a soft function and a collinear function\footnote{In some cases, e.g. the $H\to \gamma\gamma$ decay, the collinear function does not appear at next-to-leading power in the soft sector but would appear at next-to-next-to-leading power.}.
The radiative jet function depends on the light-cone component of the soft quark, denoted by $\rho$, in the form of $1/\rho$.
The integration of  $\rho$ around $0$ and $\infty$ brings about another kind of endpoint singularity\footnote{The singularity at $\rho=0$ may be avoided if the soft quark phase space is subject to a measurement function.}.

It has been realized that the two kinds of endpoint singularities are closely related to each other.
In some sense, they cancel with each other.
In order to demonstrate the cancellation clearly,
one can rearrange the integrands or construct a scaleless integral to subtract the singularity in the corresponding integration region.
Successful application can be seen, for example, in the resummation of large logarithms in $H\to \gamma\gamma$ \cite{Liu:2019oav} and ``gluon thrust'' \cite{Beneke:2022obx}.
More recently, it is discovered that the rearrangement or subtraction methods cannot be simply used to perform the resummation in the more complicated cases such as $e\mu$ backward scattering \cite{Bell:2022ott} and $B_c \to \eta_c $ decays at large recoil \cite{Bell:2024bxg}, where more general factorization pictures are revealed for the power suppressed contribution.
Therefore, it is valuable to investigate alternative methods to tame the endpoint singularities.
To this end, an appropriate regulator plays a vital role.
In this paper, we employ the analytic regulator and the $\Delta$ regulator to exhibit their features 
in dealing with the endpoint singularities.
We focus on the simple process $H\to \gamma\gamma$ and perform a region analysis of all the two-loop Feynman diagrams.
Note that the two regulators are different from that used in  ref. \cite{Liu:2019oav}.
The resulting factorization formulae and various functions would be in different forms and thus may provide complementary angles to the endpoint singularities.

The rest of the paper is organized as follows.
In section \ref{sec:QCDres}, we calculate the full result for each two-loop diagram individually. 
Then we perform a region analysis with the analytic regulator in section \ref{sec:expansion}. 
We find that no soft regions are involved.
But the cancellation of the endpoint singularities is nontrivial.
The region expansion with $\Delta$ regulators is carried out in section \ref{sec:Delta}.
Although the soft region is not vanishing in this regulator, the endpoint and ultraviolet singularities are encoded in different forms and the large logarithms can be fully reproduced only from the integration over small transverse momentum region where the leading-order (LO) results are finite.
The implication of our region analysis on the factorization and resummation is discussed in section \ref{sec:discussion}.
The conclusion is presented in section \ref{sec:conclusion}.

\section{Full one-loop and two-loop results}
\label{sec:QCDres}

\begin{figure}[H]
 \centering
{\includegraphics[width=0.4\textwidth]{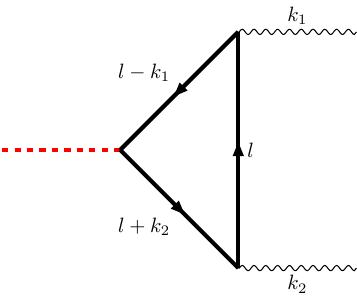}}
\caption {The one-loop diagram of the $H\to \gamma(k_1) \gamma(k_2)$ process. The diagram with the two photons exchanged is not shown.}
\label{fig:oneloop}
\end{figure}

Since the photon is massless, the $H\to \gamma\gamma$ decay is a loop-induced process; see figure \ref{fig:oneloop} for the one-loop Feynman diagram.
Although the two-loop QCD corrections with finite quark masses have been obtained in refs.\cite{Zheng:1990qa,Djouadi:1990aj,Melnikov:1993tj,Inoue:1994jq,Spira:1995rr,Fleischer:2004vb,Harlander:2005rq,Anastasiou:2006hc,Aglietti:2006tp} \footnote{The three-loop result has also been obtained numerically \cite{Niggetiedt:2020sbf}.},
the explicit analytic result corresponding to each diagram is not presented.
In this section we calculate analytically each diagram as well as the counter-term contributions independently,
which can be used to compare with the results obtained by the method of regions.

 \begin{figure}[H]
  
 \centering
 \vspace{-0.3cm}
 \subfloat[]
{\includegraphics[width=0.30\textwidth]{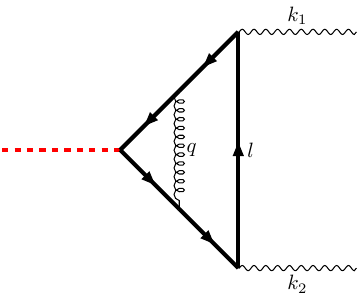}\label{fig:dia1}}
\subfloat[]
{\includegraphics[width=0.30\textwidth]{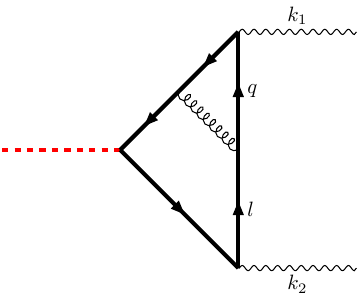}\label{fig:dia2}}
\subfloat[]
{\includegraphics[width=0.30\textwidth]{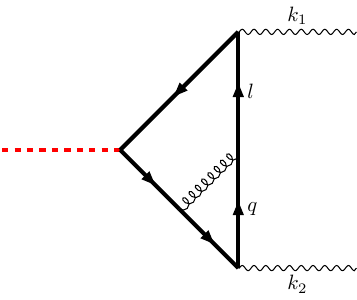}\label{fig:dia3}}

\subfloat[]
{\includegraphics[width=0.30\textwidth]{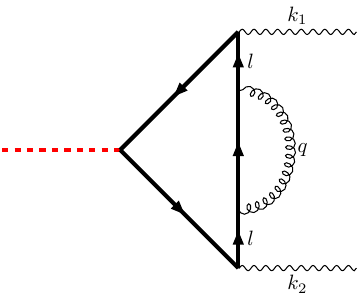}\label{fig:dia4}}
 \subfloat[]
{\includegraphics[width=0.30\textwidth]{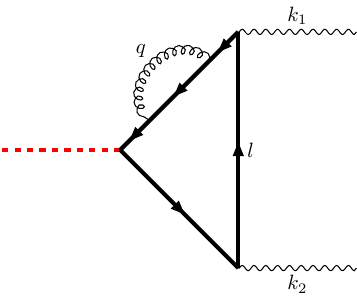}\label{fig:dia5}}
\subfloat[]
{\includegraphics[width=0.30\textwidth]{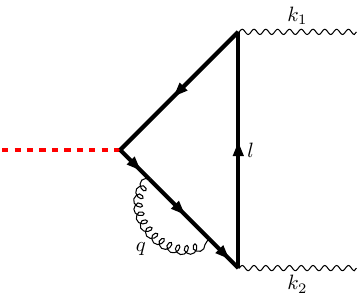}\label{fig:dia6}}
\caption{ The two-loop Feynman diagrams for the decay process $H\to \gamma(k_1) \gamma(k_2)$. The diagrams with the two photons exchanged are not shown.
}
\label{fig:two_loop_dia}
\end{figure}

We first consider the two-loop diagrams for $H\to \gamma(k_1) \gamma(k_2)$  shown in figure~\ref{fig:two_loop_dia}. 
The amplitude can be written as
\beq
\begin{aligned}
i\mathcal{M}=i \mathcal{A}^{\mu\nu} \epsilon_{\mu}^*(k_1)  \epsilon_{\nu}^*(k_2)\,.
\end{aligned}
\eeq
According to the Lorentz structure, $\mathcal{A}^{\mu\nu}$ can be decomposed into a linear combination of independent bases as
\begin{align}
     \mathcal{A}^{\mu\nu} = c_1 g^{\mu\nu} + c_2 k_1^{\mu} k_2^{\nu} + c_3 k_1^{\nu} k_2^{\mu} + c_4 k_1^{\mu} k_1^{\nu} + c_5 k_2^{\mu} k_2^{\nu}\,.
\end{align}
The Ward identity indicates that $k_{1\mu}A^{\mu\nu}=k_{2\nu}A^{\mu\nu}=0$, which leads to 
\begin{align}
 \mathcal{A}^{\mu\nu} = c_1 \left( g^{\mu \nu}- \frac{k_1^{\nu} k_2^{\mu}}{k_1\cdot k_2} \right)+ c_2 k_1^{\mu} k_2^{\nu} \,,
\end{align}
where the last term does not contribute to the amplitude $\mathcal{M}$ due to $k_1\cdot \epsilon^*(k_1)=0$.
The coefficient $c_1$ can be projected out by
\begin{align}
    c_1 = g^{\rho\sigma} \left( -g_{\rho\mu} + \frac{k_{1\rho} k_{2\mu}+k_{1\mu} k_{2\rho}}{k_1\cdot k_2}  \right)
     \left( -g_{\sigma\nu} + \frac{k_{1\sigma} k_{2\nu}+k_{1\nu} k_{2\sigma}}{k_1\cdot k_2}  \right)  \mathcal{A}^{\mu\nu}\,.
     \label{eq:projector}
\end{align}
Defining $g_{\perp}^{\mu\nu} = g^{\mu\nu} - ( k_1^{\nu} k_2^{\mu} + k_1^{\mu} k_2^{\nu} )/(k_1\cdot k_2)$, we can write
\begin{align}
    i\mathcal{M}= i c_1 g_{\perp}^{\mu\nu} \epsilon_{\mu}^*(k_1)  \epsilon_{\nu}^*(k_2) 
    = i c_1  \epsilon_{\perp}^*(k_1) \cdot \epsilon_{\perp}^*(k_2)\,.
\end{align}

In this paper, we are going to calculate $c_1$ up to two-loop level, i.e.,
\begin{align}
c_1 = \frac{ N_c e_b^2\alpha y_b m_b} {2\pi}
\left( \mathcal{A}^{(0)}+ \frac{\alpha_s C_F}{4\pi} \mathcal{A}^{(1)}   \right)\,,
\label{eq:c1res}
\end{align}
where $N_c$ is the number of quark colors, $e_b$ is the electric charge of the bottom quark, $\alpha$ is the electromagnetic coupling, $y_b$ and $m_b$ are the bottom quark Yukawa coupling and mass, respectively.
We have expanded $c_1$ perturbatively in terms of $\alpha_s C_F/4\pi$ with $\alpha_s$ being the strong coupling and $C_F$ being the Casimir operator of the $SU(N_c)$ group.
The LO result can be obtained by calculating the one-loop diagram shown in figure \ref{fig:oneloop}, and is given by
\beq
\begin{aligned}
\mathcal{A}^{(0)}=(1-4R) \ln ^2\omega - 4 = \ln^2(-R)-4-4R\left( \ln^2(-R)-\ln(-R)   \right)+\mathcal{O}(R^2)
\label{eq:LOA}
\end{aligned}
\eeq
with $R=m_b^2/m_H^2-i0^+$ and 
\begin{align}
\omega=\frac{\sqrt{1-4R}-1}{\sqrt{1-4R}+1}+i0^+ \,.
\end{align}
Here we have written the imaginary part explicitly in order to fix the continuation direction when needed.
The contribution from diagrams with the two photons exchanged has already been included in $c_1$.

The two-loop corrections consist of the Feynman diagrams shown in figure \ref{fig:two_loop_dia}.
After applying the projector in eq. (\ref{eq:projector}), we obtain the result of  $\mathcal{A}^{(1)}$ in terms of scalar integrals.
These scalar integrals are not independent and have been reduced to a set of basis integrals, called master integrals, using the {\tt FIRE} \cite{Smirnov:2019qkx} package which implements the Laporta's algorithm \cite{Laporta:2000dsw}.
The topologies of these master integrals are depicted in figure \ref{fig:topo}.
Five of the two-loop master integrals are products of one-loop integrals, and thus can be obtained simply.
The others were calculated using the method of differential equations \cite{Kotikov:1990kg,Kotikov:1991pm}.
In particular, we have transformed the differential equations to the canonical form \cite{Henn:2013pwa} using the {\tt Libra} package \cite{Lee:2020zfb}.
The boundary condition is set at the limit of $m_b\to \infty$.
In this limit, all the master integrals are reduced to vacuum massive integrals, which have been calculated analytically in ref. \cite{Steinhauser:2000ry}.
Then the solutions can be derived and expressed in terms of multiple polylogarithms (MPLs) \cite{Goncharov:1998kja}
defined by
\begin{align}
    G(a_1,a_2,...,a_n,x)\equiv \int_0^x \frac{dt}{t-a_1}G(a_2,...,a_n,t)
\end{align}
and
\begin{align}
    G(\vec{0}_n,x)\equiv \frac{1}{n!}\ln^n x \,.
\end{align}
We work in $D=4-2\epsilon$ dimensional spacetime,
and the complete results for the two-loop diagrams are collected in appendix \ref{sec:app1}.
Here we only present the expansions up to $\mathcal{O}(R)$:
\begin{align}
    \mathcal{A}^{(1)a} =&\left( \frac{\mu^2}{m_H^2}\right)^{2\epsilon} \left[  \frac{1}{\epsilon}\left(   4 \ln ^2(-R)-16    \right)  -\frac{1}{6}\ln^4(-R)+4\ln^3(-R)-8 \ln ^2(-R) \ln R \right.\nn
    \\&\left.  +\left(6-\frac{4\pi^2}{3}\right)\ln^2(-R)    - \left(32\zeta(3)+\frac{4\pi^2}{3}+32\right)\ln(-R)+32\ln R-56\zeta(3) \right.\nn
    \\&\left. +\frac{10\pi^2}{3} -\frac{2\pi^4}{5}-94  \right.\nn
    \\&\left.    +R\left(\frac{1}{\epsilon}\left( -16 \ln ^2(-R) +16 \ln (-R) \right) + \frac{2}{3} \ln ^4(-R)-\frac{64}{3} \ln ^3(-R)+32\ln^2(-R)\ln R \right.\right.\nn
    \\&\left.\left.  + \left(\frac{20 \pi ^2}{3}-36\right) \ln ^2(-R)-32\ln(-R)\ln R+  \left(128 \zeta (3)-\frac{8 \pi ^2}{3}+96\right) \ln (-R)  \right.\right.\nn
    \\&\left.\left.   +64 \zeta (3)+\frac{92 \pi ^4}{45}-\frac{56 \pi ^2}{3}-200  \right)    \right]+\mathcal{O}(R^2) \,, 
    \label{eq:expand_NLOa}
    \\
    \mathcal{A}^{(1)b} =&\left( \frac{\mu^2}{m_H^2}\right)^{2\epsilon} \left[  \frac{1}{\epsilon}\left(  \ln ^2(-R)-4    \right) + \frac{1}{3} \ln ^3(-R)-2 \ln ^2(-R) \log R+8 \ln R\right.\nn
    \\&\left.-\left(\pi ^2 +8 \right) \ln (-R)-6 \zeta (3)-2 \pi ^2-22\right.\nn
    \\&\left. +R\left(\frac{1}{\epsilon}\Big( -4 \ln ^2(-R)+4 \ln (-R) \Big) -\frac{2}{3} \ln ^3(-R)+8 \ln ^2(-R) \ln R\right.\right.\nn
    \\&\left.\left. +\left(-\frac{2 \pi ^2}{3}+2\right) \ln ^2(-R)-8 \ln (-R) \ln R    +\left(\frac{28 \pi ^2}{3}  +48\right) \ln (-R)     \right.\right.\nn
    \\&\left.\left.  +136 \zeta (3)-\frac{2 \pi ^4}{9}+\frac{34 \pi ^2}{3}-16 \right)  \right]+\mathcal{O}(R^2)\,,
    \label{eq:expand_NLOb}
    \\
    \mathcal{A}^{(1)d} =&\left( \frac{\mu^2}{m_H^2}\right)^{2\epsilon} \Bigg[  -\frac{6}{\epsilon^2}+ \frac{1}{\epsilon}\Big( 2 \ln ^2(-R)+12 \ln (-R)+12\ln R+23     \Big)  \nn
    \\&  + \frac{4}{3} \ln ^3(-R)-4 \ln ^2(-R) \ln R+4 \ln ^2(-R)-24 \ln (-R) \ln R-12 \ln ^2R \nn
    \\& -46 \ln R+\left(22-\frac{2 \pi ^2}{3}\right) \ln (-R)-20 \zeta (3)+\frac{81}{2}-\frac{7 \pi ^2}{3}   \nn
    \\& -R\bigg(\frac{1}{\epsilon}\Big( 20 \ln ^2(-R)+112 \ln (-R)+168 \Big)+ \frac{40}{3} \ln ^3(-R)-40 \ln ^2(-R) \ln R   \nn
    \\&+108 \ln ^2(-R)-224 \ln (-R) \ln R +\left(368+\frac{4 \pi ^2}{3}\right) \ln (-R)-336 \ln R   \nn
    \\& -56 \zeta (3)-8 \pi ^2+696 \bigg)  \Bigg]+\mathcal{O}(R^2) \,,
    \label{eq:expand_NLOd}
    \\
    \mathcal{A}^{(1)e} =&\left( \frac{\mu^2}{m_H^2}\right)^{2\epsilon} \left[  \frac{3}{\epsilon^2}  -  \frac{1}{\epsilon}\left(  \ln ^2(-R)+6\ln R+\frac{23}{2}    \right)    -\ln ^3(-R)+2 \ln ^2(-R) \ln R\right.\nn
    \\&\left.-\ln ^2(-R)+6 \ln ^2R+\left(\frac{\pi ^2}{3}-3\right) \ln (-R)+23 \ln R +30 \zeta (3)+\frac{\pi^2}{2}-\frac{93}{4}  \right.\nn
    \\&\left.    +R\left(\frac{1}{\epsilon}\Big( -2 \ln ^2(-R)+44 \ln (-R)+96 \Big) -2 \ln ^3(-R)+4 \ln ^2(-R) \ln R\right.\right.\nn
    \\&\left.\left.+36 \ln ^2(-R)-88 \ln (-R) \ln R+ \left(152+\frac{2 \pi ^2}{3}\right) \ln (-R)-192 \ln R\right.\right.\nn
    \\&\left.\left.  -36 \zeta (3)-\frac{22 \pi ^2}{3}+356 \right)  \right]+\mathcal{O}(R^2)\,,
    \label{eq:expand_NLOe}
\end{align}
where $\mu$ denotes the renormalization scale and $\zeta(n)$ is the Riemann zeta function.
The results of diagrams $(c)$ and $(f)$ are the same as $\mathcal{A}^{(1)b}$ and  $\mathcal{A}^{(1)e}$, respectively.
We find that $\mathcal{A}^{(1)a}$ contains $\ln^4 (-R)$ while
the other diagrams have at most $\ln^3 (-R)$ logarithmic terms at LP\footnote{Due to the suppression factor $y_bm_b$ in the coefficient, the LP expansion of $\mathcal{A}^{(1)}$ corresponds to $\mathcal{O}(R)$ of the whole amplitude, which is often called a NLP contribution; 
see eq. (\ref{eq:c1res}).}.
The power suppressed terms, i.e., the $\mathcal{O}(R)$ terms, exhibit similar logarithmic structures.

The above two-loop results still contain ultraviolet divergences in the form of $1/\epsilon$ poles\footnote{There are double poles $1/\epsilon^2$ in the results of some two-loop diagrams; see $\mathcal{A}^{(1)d}$ and $\mathcal{A}^{(1)e}$ in eq. (\ref{eq:expand_NLOd}) and eq. (\ref{eq:expand_NLOe}), respectively. They are guaranteed to cancel due to gauge invariance, which is confirmed by explicit calculation.} that have to be canceled after including the contribution from mass renormalization.
We have chosen on-shell renormalization scheme for both the quark mass and the Yukawa coupling.
After obtaining the renormalization constant, one can insert the counter-term vertices into the one-loop diagrams and calculate the loop integrals up to finite terms in $\epsilon$.
Instead, we can compute the contribution by substituting $m_b$ by $ Z_m m_b$ in eq. (\ref{eq:LOA}) (with higher orders in $\epsilon$), where  $Z_m$ is the renormalization constant for the bottom quark mass.
The renormalized two-loop correction is shown in appendix \ref{sec:app1},
and its expansion in $R$ is given by
\begin{align}
    \mathcal{A}^{(1)} =& -\frac{1}{6}\ln^4(-R)+2 \ln^3(-R)-\frac{4\pi ^2}{3}  \ln ^2(-R)-\ln(-R)\left(   32 \zeta (3)+\frac{4 \pi ^2}{3}+24   \right) \nn
    \\& +8 \zeta (3)-\frac{2 \pi ^4}{5}-40       \nn
    \\& +R\left( \frac {2} {3}\ln^4 (-R) - 
 24\ln^3 (-R) + \ln^2 (-R)\left (20 + \frac {16\pi^2} {3} \right)\right. \nn
 \\&\left.  +(128\zeta (3) +32)\ln (-R) + 
 32\zeta (3) + \frac {8\pi^4} {5} - \frac {8\pi^2} {3} - 128 \right)+\mathcal{O}(R^2)\,.
 \label{eq:nloA}
\end{align}
The above analytic results serve as standard references to compare with for the predictions from the method of regions or an effective field theory.

\section{Expansion by regions and analytic regulators}
\label{sec:expansion}

It is intriguing to study the logarithmic structure of the above results in eq. (\ref{eq:LOA}) and eq. (\ref{eq:nloA}).
The logarithms appear due to the presence of multiple scales in the process.
To see the contribution from individual scales, 
we employ the strategy of expansion by regions \cite{Smirnov:1997gx,Beneke:1997zp}.
The power counting parameter in our problem is chosen as $\lambda\sim \sqrt{R}$,
which is considered to be much less than one.
The power counting for a momentum is conveniently carried out on its light-cone components, which are defined via
\beq
\begin{aligned}
p^{\mu}&=n\cdot p\frac{\bar n^{\mu}}{2}+\bar n\cdot p\frac{n^{\mu}}{2}+p^{\mu}_{\bot}\\&
=(n\cdot p,\bar n\cdot p,p^{\mu}_{\bot}).
\end{aligned}
\eeq
Here $n$ and $\bar{n}$ are two light-cone vectors
satisfying $n\cdot \bar{n}=2$.
Specifically, we set $n^{\mu}= k_1^{\mu}/k_1^0$ and $\bar{n}^{\mu}= k_2^{\mu}/k_2^0$ in our case.
We will need the following momentum modes in our analysis,
\beq
\begin{aligned}
 \text{hard}~(h):\ \ & l^{\mu}\sim (1, 1 ,1)m_H \,,
 \\ \text{collinear}~(c):\ \ & l^{\mu}\sim (\lambda^2, 1 ,\lambda)m_H\,,
 \\ \text{anti-collinear}~(\bar c):\ \ & l^{\mu}\sim (1, \lambda^2 ,\lambda)m_H \,, 
 \\ \text{soft}~(s):\ \ & l^{\mu}\sim (\lambda, \lambda ,\lambda)m_H \,.
\end{aligned}
\label{eq:modes1}
\eeq
When there is a propagator that splits into two particles of a collinear and a soft momentum modes, respectively, 
we also need the hard-collinear and hard-anti-collinear modes:
\beq
\begin{aligned}
\text{hard-collinear}~(hc):\ \ & l^{\mu}\sim (\lambda, 1 ,\lambda^{1/2})m_H\,,
 \\ \text{hard-anti-collinear}~(h\bar c):\ \ & l^{\mu}\sim (1, \lambda ,\lambda^{1/2})m_H \,.
\end{aligned}
\label{eq:modes2}
\eeq

Then the next step is to determine all the regions that provide non-vanishing contributions.
The identification of the regions was usually made using heuristic methods based on experience in momentum space.
In some cases, momentum shifts are required to find out all the regions \cite{terHoeve:2023ehm}.
A more systematical method is developed from the geometric point of view, i.e., the regions are determined as lower facets of the corresponding Newton polytope.
This method is based on the Feynman parameter representation of the loop integrals\footnote{The Lee-Pomeransky representation of loop integrals would generate the same regions \cite{Gardi:2022khw}.}.
Using the parameters $x_1,x_2,...,x_n$ corresponding to the denominators of propagators,
the scaling of the parameters are indicated by $x_i \sim \lambda^{v_i}$.
The relevant regions are specified by the multi-dimensional vectors $\vec{v}_i=(v_{i,1},v_{i,2},...,v_{i,n})$.
The algorithms to determine these region vectors have been proposed in refs. \cite{Pak:2010pt,Ananthanarayan:2018tog,Heinrich:2021dbf},
and one can make use of the packages {\tt Asy2.1} \cite{Jantzen:2012mw} and {\tt pySecDec} \cite{Heinrich:2021dbf} to find the regions automatically\footnote{Recently, hidden regions are revealed at three-loop level, which arise due to the cancellation between  terms of opposite signs in the Lee-Pomeransky polynomials \cite{Gardi:2024axt}.  }.
Note that the region vectors are equivalent if they are different by a vector $(1,1,\cdots,1)a$ with $a$ being an arbitrary constant.

The regions discovered using the above algorithms are closely related to the solutions of the Landau equations,
which have a clear interpretation in momentum space \cite{Ananthanarayan:2018tog,Heinrich:2021dbf}.
The endpoint singularities derived from the Landau equations\footnote{Here we have abused the terminology ``endpoint singularities''. It is understood to be different from those mentioned in the introduction. } correspond to the hard propagators, 
and the other propagators can be classified into collinear and soft sub-graphs.
However, the solutions of the Landau equations can be considered as regions only if they satisfy some requirements; see ref. \cite{Gardi:2022khw} for details.
In our work, we take a more intuitive method.
Firstly, the scaling of the virtuality of each propagator $l_i^2-m_i^2$ is determined by the inverse of the scaling of the corresponding parameter $x_i$ \cite{Gardi:2022khw},
i.e., $l_i^2-m_i^2 \sim \lambda^{-v_i}$.
Then the scaling of the light-cone components of each momentum is fixed after applying momentum conservation at all vertices.
In most cases the solutions are unique when the momentum flow connects to the external particles.

The expansion by regions is often conducted in dimensional regularization for the infrared and ultraviolet divergences.
This regulator has the advantage that the scaleless integrals can be considered vanishing.
However, it is not sufficient to regulate all divergences with such a regulator.
The endpoint divergence we are going to show below is an example.
We will adopt two kinds of regulators to calculate the integrals containing such divergences.
In the first method, we increase the powers of the propagators from 1 to $1+\eta$ with $\eta$ being a complex number, called {\it the analytic regulator}, so that 
the endpoint divergence is encoded to a $1/\eta$ pole.
Since the Symanzik polynomials are not changed, no new regions appear.
In the second method, we add a small mass scale $\Delta$ to the denominators that cause endpoint divergences.
The resulting divergence is cast in the form of $\ln \Delta$.
Additional regions may appear in this regulator.
We will discuss these two methods in turn.

\subsection{One-loop amplitudes}

\begin{figure}[t]
 \centering
 \vspace{0cm}
 \subfloat[]
{\includegraphics[width=0.3\textwidth]{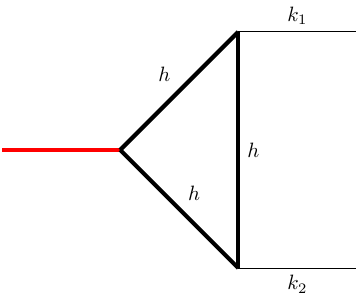}}
\subfloat[]
{\includegraphics[width=0.3\textwidth]{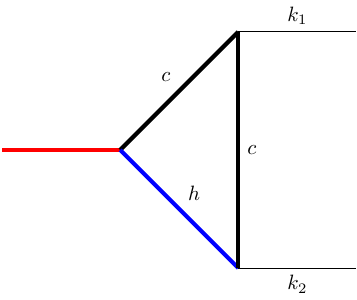}}
 \subfloat[]
{\includegraphics[width=0.3\textwidth]{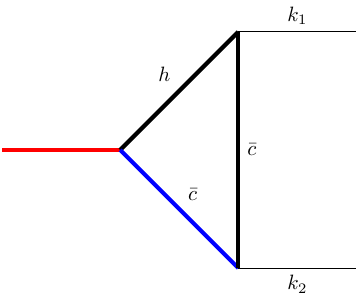}}
\caption {Three regions in the one-loop amplitude. 
The momentum scaling mode of each propagator is shown explicitly by the labels, i.e., $h$, $c$ and $\bar{c}$. 
An analytic regulator is imposed on the blue line.}
\label{fig:oneloop_region}
\end{figure}

Let us start with the LO calculation.
From the Feynman diagram shown in figure \ref{fig:oneloop}, we can write \footnote{Here we did not include the contribution from the diagram with two photons exchanged.}
\begin{align}
\mathcal{A}^{(0)}= & \frac{32i\pi^2 {\tilde{\mu}}^{2\epsilon}  }{(D-2)m_H^2} \int\frac{d^Dl}{(2\pi)^D} \frac{1}{(-l^2+m_b^2-i0)[-(l-k_1)^2+m_b^2-i0]}\nn
\\& \frac{1}{[-(l+k_2)^2+m_b^2-i0]}\big[ (2D-12)m_H^2l^2-(D-2)(2m_b^2m_H^2-m_H^4)\nn
\\& +32(k_1\cdot l)(k_2\cdot l)  \big]
\label{eq:loA}
\end{align}
with $\tilde{\mu}^2 = \mu^2 / (4\pi e^{-\gamma_E}) $.
We find that the loop momentum $l$ can be hard, collinear and anti-collinear when using expansion by regions.
The three regions are shown in figure \ref{fig:oneloop_region}.

In the hard region, we can simply drop the small scale $m_b$ in both the numerator and denominator, and then the corresponding  contribution is given by
\begin{align}
\mathcal{A}^{(0)}_{h} & = \frac{32i\pi^2 {\tilde{\mu}}^{2\epsilon}}{(D-2)m_H^2} \int\frac{d^Dl}{(2\pi)^D}\frac{ (2D-12)m_H^2l^2+(D-2)m_H^4+32(k_1\cdot l)(k_2\cdot l)}{(-l^2-i0)[-(l-k_1)^2-i0][-(l+k_2)^2-i0]} \nn\\
&=  \left( -\frac{\mu^2}{m_H^2} \right)^{\epsilon}  
\left[  \frac{2}{\epsilon ^2}-4-\frac{ \pi ^2}{6}  +\mathcal{O}(\epsilon)    \right]\,.
\label{eq:one_loop_hard_region_eta}
\end{align}
It is clear that the above result does not have large logarithms at the hard scale $m_H$.

The contribution from the collinear region reads
\begin{align}
\mathcal{A}^{(0)}_{c}=  32i\pi^2m_H^2  \int\frac{d^Dl}{(2\pi)^D}\frac{(\tilde{\mu}^2)^\epsilon(\nu^2)^\eta}{(-l^2+m_b^2-i0)[-(l-k_1)^2+m_b^2-i0][-m_H(\bar n\cdot l)-i0]^{1+\eta}}.
\end{align}
Notice that the third propagator becomes proportional to $1/\bar{n}\cdot l$ after expansion.
After writing $d^Dl = 1/2 d(n\cdot l) d (\bar{n}\cdot l)  d^{D-2} l_{\perp}$, we can perform the integration over $n\cdot l$ using the residue theorem. 
The non-vanishing contribution requires $0 < \bar{n}\cdot l < m_H$, which leads to a divergence in the integration of $\bar{n}\cdot l$ near the endpoint at 0.
To regulate this divergence, we have increased the power of the propagator and add a factor $(\nu^2)^{\eta} $ to balance the mass dimension.
This kind of regulator is called an analytic regulator.
With the analytic regulator, the above integral can be evaluated as
\begin{align}
\mathcal{A}^{(0)}_{c} & =  -32\pi^3m_H  \int\frac{d^{D-2}l_{\perp}}{(2\pi)^D}\int_0^{m_H} d (\bar{n}\cdot l)\frac{(\tilde{\mu}^2)^\epsilon(\nu^2)^\eta}{(-l_{\perp}^2+m_b^2)[-m_H(\bar n\cdot l)]^{1+\eta}} \nn\\
 &=  -32\pi^3 \left( -\frac{\nu^2}{m_H^2}\right)^{\eta}\frac{1}{\eta} \int\frac{d^{D-2}l_{\perp}}{(2\pi)^D}\frac{(\tilde{\mu}^2)^\epsilon}{(-l_{\perp}^2+m_b^2)}\nn\\
&=\left( \frac{\mu^2}{m_b^2} \right)^{\epsilon} \left( -\frac{\nu^2}{m_H^2}\right)^{\eta} 
\left[  -\frac{2}{\eta \epsilon} + \mathcal{O}(\eta)+\mathcal{O}(\epsilon)        \right]\,,
\label{eq:AcLO}
\end{align}
where the $\eta$ and $\epsilon$ poles correspond to the endpoint and ultraviolet divergences, respectively.

The contribution from the anti-collinear region reads
\begin{align}
\mathcal{A}^{(0)}_{\bar c}=   32i\pi^2m_H^2 \int\frac{d^Dl}{(2\pi)^D}\frac{(\tilde{\mu}^2)^\epsilon(\nu^2)^\eta}{(-l^2+m_b^2-i0)[m_H(n\cdot l)-i0][-(l+k_2)^2+m_b^2-i0]^{1+\eta}} \, .
\end{align}
We have increased the power of the same propagator as in the collinear region.
But the analytical regulator is imposed on an anti-collinear propagator now.
After integration over $\bar{n}\cdot l$ with the residue theorem, a factor $(n\cdot l)^{\eta}$ arises
so that the endpoint divergence at $n\cdot l =0$ is regulated.
Meanwhile, the dimension of the  perpendicular integrand is also changed by $\eta$ 
and thus the ultraviolet divergence also depends on $\eta$, for example, in the form of $1/(\eta+\epsilon)$.
Since the ultraviolet divergences in the other regions are regulated only in $\epsilon$, we have to expand the result around $\eta=0$ before expansion in $\epsilon$ for consistency.     
Otherwise, the poles of $\eta$ and $\epsilon$ cannot be canceled in the sum of all relevant regions.
This treatment is also in line with our motivation of introducing $\eta$ only for the endpoint divergences.
The result of the above equation can be obtained by 
\begin{align}
\mathcal{A}^{(0)}_{\bar c} & =  32\pi^3  \int\frac{d^{D-2}l_{\perp}}{(2\pi)^D}\int_{-m_H}^{0} d (n\cdot l)\frac{(\tilde{\mu}^2)^\epsilon (\nu^2/m_H)^\eta }{(-l_{\perp}^2+m_b^2)^{1+\eta}[-( n\cdot l)]^{1-\eta}} \nn\\
& =     \left( \frac{4\pi\tilde{\mu}^2}{m_b^2} \right)^{\epsilon} \left(  \frac{\nu^2}{m_b^2}\right)^{\eta}\frac{\Gamma(\epsilon+\eta)}{\eta\Gamma(1+\eta)}\nn\\
& =  \left( \frac{\mu^2}{m_b^2} \right)^{\epsilon} \left(  \frac{\nu^2}{m_b^2}\right)^{\eta} 
\left[  \frac{2}{\eta \epsilon}   - \frac {2} {\epsilon^2}  + \frac {\pi^2} {6}   + \mathcal{O}(\eta)+\mathcal{O}(\epsilon)       \right]\,,
\label{eq:AcbarLO}
\end{align}
where the first term in the bracket is the same as that in the collinear case but with an opposite sign
and the double pole $1/\epsilon^2$ results from the expansion of $\Gamma(\epsilon+\eta)$ in $\eta$.
We see from eq. (\ref{eq:AcLO}) and eq. (\ref{eq:AcbarLO}) that the natural collinear and anti-collinear scales are $\mu\sim m_b$.
But the natural scales of the endpoint logarithms are $m_H$ and $m_b$ for the collinear and anti-collinear contributions, respectively.
This can be understood because the propagator with an $\eta$ regulator is of order $\mathcal{O}(1)$ and $\mathcal{O}(\lambda^2)$ in the two different regions, respectively. 
The presence of an $\eta$  regulator in eq. (\ref{eq:AcLO}) actually breaks the symmetry that the integral is invariant under the rescaling of $\bar{n}\to e^{-\alpha} \bar{n}$ \footnote{The anti-collinear integral with an $\eta$ regulator breaks the rescaling symmetry under  $n\to e^{\alpha} n $.},
which is an effect called collinear anomaly \cite{Becher:2010tm}. 
As a consequence, the sum of the collinear and anti-collinear regions produces a logarithm which involves a hard scale,
\begin{align}
    \left( \frac{\nu^2}{-m_H^2}\right)^{\eta} \frac{1}{\eta}
    +\left(  \frac{\nu^2}{m_b^2}\right)^{\eta} \frac{-1}{\eta}=\ln (-R)\,.
\end{align}
We emphasize that one could have added the $\eta$ regulator on the other propagator, e.g., the second propagator in eq. (\ref{eq:loA}).
Then the results of $\mathcal{A}^{(0)}_{c}$  and $\mathcal{A}^{(0)}_{\bar c}$ would be exchanged. 
However, their sum remains the same, independent of the way to impose the $\eta$ regulator.

Summarizing the above results in different regions, we obtain
\begin{align}
\mathcal{A}^{(0)}_{\rm exp}\equiv \mathcal{A}^{(0)}_{h}+ \mathcal{A}^{(0)}_{c}+
\mathcal{A}^{(0)}_{\bar c}
=\ln^2 (-R) - 4\,,
\label{eq:Aexp}
\end{align}
where both the $\eta$ and $\epsilon$ poles are canceled out.
The above sum agrees with the full result in eq. (\ref{eq:LOA}) at LP.

\subsection{Two-loop amplitudes}

The next-to-leading order corrections consist of the one-loop amplitudes with counter-term vertices 
and the two-loop amplitudes.
The former can be calculated with the same procedure described in the above subsection.
Below we discuss only the calculation of two-loop amplitudes.
The diagrams \ref{fig:two_loop_dia}(c) and \ref{fig:two_loop_dia}(f) give the same contributions to $c_1$ as \ref{fig:two_loop_dia}(b) and \ref{fig:two_loop_dia}(e), respectively.
Therefore, we only focus on the diagrams \ref{fig:two_loop_dia}(a), (b), (d), (e).

In the standard method, the loop amplitudes are often reduced to a set of basis integrals, called master integrals, by using the integration by parts identities firstly \cite{Tkachov:1981wb,Chetyrkin:1981qh} so that only a small number of integrals need to be calculated.
But the reduction coefficients of the master integrals may contain inverse powers of $m_b^2$.
In this case, the master integrals must be expanded to higher orders of $m_b^2$.
To avoid such complexities, we choose to expand the whole loop amplitude directly.

\subsubsection{Diagram \ref{fig:two_loop_dia}(a)}

\begin{figure}[h]
 \centering
 \vspace{0cm}
 \subfloat[ ]
{\includegraphics[width=0.3\textwidth]{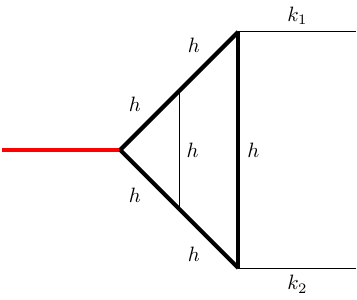}}
\subfloat[]
{\includegraphics[width=0.3\textwidth]{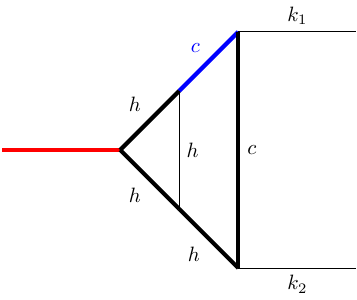}}
 \subfloat[]
{\includegraphics[width=0.3\textwidth]{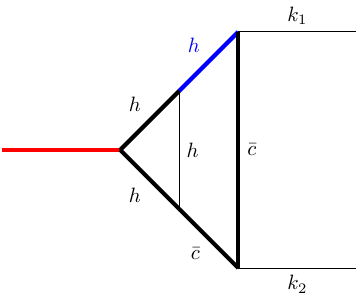}}

 \subfloat[]
{\includegraphics[width=0.3\textwidth]{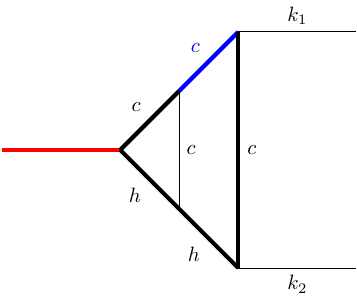}}
\subfloat[]
{\includegraphics[width=0.3\textwidth]{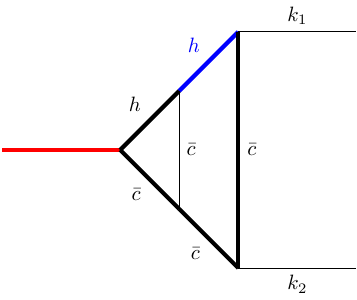}}
\caption {Five regions at LP for the diagram~\ref{fig:two_loop_dia}(a). 
The momentum scaling mode of each propagator is shown explicitly by the labels ($c$, $\bar{c}$ and $h$). An analytic regulator is imposed on the blue line. }
\label{fig:dia_a_eta}
\end{figure}

When calculating the diagram \ref{fig:two_loop_dia}(a) with the method described above,
we find five regions that give the LP contributions,
which are shown in figure~\ref{fig:dia_a_eta} with explicit scaling mode labels on the propagators.

In the hard region shown in figure \ref{fig:dia_a_eta}(a), the loop momenta are considered of order $m_H$ and the quark mass is neglected. 
After reduction to master integrals and integration over Feynman parameters, the two-loop amplitude is evaluated to
\begin{align}
\mathcal{A}^{(1)a}_{(h-h)}=&  \left( -\frac{\mu^2}{m_H^2} \right)^{2\epsilon}  
\left[-\frac{1}{\epsilon ^4}  +\frac{4}{\epsilon ^3}+\frac{1}{\epsilon ^2}\left(    -\frac{5 \pi ^2}{6} +4 \right) -\frac{1}{\epsilon }\left(   \frac{58\zeta (3)}{3} +\frac{2 \pi ^2}{3}+8\right)  \right.\nn
\\&\left.        -\frac{128\zeta (3)}{3}-\frac{3 \pi ^4}{8}+\frac{4 \pi ^2}{3}-78      \right]\,,
\label{eq:A1ahh}
\end{align}
where the subscript $(h-h)$ refers to the  scaling of the 
loop momenta $l$ and $q$ in figure \ref{fig:two_loop_dia}, i.e., $l$ is hard (denoted by the first $h$) and $q$ is hard (denoted by the second $h$), too.
No logarithms will appear if the scale $\mu$ is chosen at $m_H$. 

In the other regions, endpoint singularities require an additional regulator.
As in the one-loop integrals, we improve the power of one specific propagator to $1+\eta$.
Such a propagator is denoted by a blue line in figure \ref{fig:dia_a_eta}.
We will take the $(c-h)$ region as an example to show the details in our calculation.
In this region, we need to calculate the following integral
\begin{align}
I_{(c-h)}&=\int \frac{d^Dl}{(2\pi)^D} \frac{d^Dq}{(2\pi)^D} \frac{1}{(-l^2+m_b^2-i0)[-l^2+m_H(n\cdot l)+m_b^2-i0]^{1+\eta}[-m_H(\bar n\cdot l)-i0]} \nn
\\
&\times \frac{1}{(-q^2-i0)[-q^2+m_H(n\cdot q)-(\bar n\cdot l)(n\cdot q)-i0]} \nn
\\
&\times \frac{1}{[-q^2-m_H(\bar n\cdot q)-(\bar n\cdot l)(n\cdot q)-m_H(\bar n\cdot l)-i0]}.
\label{eq:two_loop_example}
\end{align}
The presence of $(\bar n\cdot l)( n\cdot q)$ in the denominator prevents usage of automatic packages in IBP reductions
and the calculation using  Feynman parameters\footnote{The same problem was encountered in ref. \cite{terHoeve:2023ehm}, where the authors chose to perform the IBP reduction loop by loop.}.
Therefore, we retain the full form of the propagators
and perform the expansion in the Feynman parameter space, obtaining
\begin{align}
I_{(c-h)}=&-\frac{16^{\epsilon -2} \pi ^{2 \epsilon -4} \Gamma (\eta +2 \epsilon +2)}{\Gamma (\eta +1)}\int\left( \prod_{i=1}^{6} dx_i\right) \delta\left(\sum_{i=1}^{6} x_i -1\right)x_2^{\eta }    \nn
\\&\times [(x_1+x_2) (x_4+x_5+x_6)]^{\eta +3 \epsilon} \nn
\\& \times \left[m_b^2 (x_1+x_2)^2 (x_4+x_5+x_6)
\right.\nn\\&\left.
-m_H^2 (x_1 x_5 x_6+x_2 (x_3 (x_4+x_5+x_6)+x_6 (x_4+x_5)))\right]^{-\eta -2 \epsilon -2}\,, 
\end{align}
which corresponds to the region vector 
$(-2,-2,0,0,0,0)$.
Here, we do not need to expand the argument of the $\delta$-function because of the Cheng-Wu theorem \cite{Cheng:1987ga}. 
For example, we can replace the $\delta$-function $\delta\left(\sum_{i=1}^{6} x_i -1\right)$ by $\delta\left( x_4+x_5 -1\right)$. 
Then we find it convenient to introduce the variables $z$ and $w$ by defining $x_1=zw, x_2=z(1-w)$ to simplify the subsequent calculation. 
After finishing all the integrals, we obtain the result
\begin{align}
I_{(c-h)}=\frac{4^{3 \epsilon -4} \pi ^{2 \epsilon -\frac{5}{2}} \left(-m_H^2\right)^{-\epsilon } m_b^{-2 (\eta +\epsilon )} \csc (\pi  \epsilon ) \Gamma (\epsilon +\eta ) [\psi _{0}(\eta )-\psi _{0}(\eta -\epsilon )]}{m_H^4 \epsilon  \Gamma (\eta +1) \Gamma \left(\frac{1}{2}-\epsilon \right)},
\end{align}
where $\psi _{0}(x)$ is the digamma function defined as $\psi_0(x)=d \ln \Gamma (x) / d x $.
The two-loop amplitude in this region is given by
\begin{align}
\mathcal{A}^{(1)a}_{( c- h)}=&\left( -\frac{\mu^2}{m_H^2} \right)^{\epsilon}\left( \frac{\mu^2}{m_b^2} \right)^{\epsilon} \left( \frac{\nu^2}{m_b^2} \right)^{\eta} \left[  \frac{1}{\eta}\left(  -\frac{4}{\epsilon ^3}-\frac{4}{\epsilon } +\frac{32\zeta (3)}{3} -8\right) \right.\nn
\\&\left.   -\frac{4}{\epsilon ^3}-\frac{4}{\epsilon ^2}    +    \frac{1}{\epsilon}\Big(   8 \zeta (3)-8  \Big)    +\frac{32 \zeta (3)}{3}-16   \right]\,.
\end{align}  
The endpoint singularity is still in the form of $1/\eta$ while the infrared and ultraviolet divergences manifest themselves as poles up to $1/\epsilon^3$.
The factors in front of the bracket embody the scales of the loop momenta and the propagator with an $\eta$ regulator.

The results of the other regions can be calculated by the same method, and are given by
\begin{align}
\mathcal{A}^{(1)a}_{( \bar c- h)}=&\left(  -\frac{\mu^2}{m_H^2}    \right)^{\epsilon}\left(  \frac{\mu^2}{m_b^2}  \right)^{\epsilon}\left(  -\frac{\nu^2}{m_H^2}  \right)^{\eta} \left[ \frac{1}{\eta}\left( \frac{4}{\epsilon^3}+\frac{4}{\epsilon} -\frac{32 \zeta (3)}{3}+8 \right) -\frac{4}{\epsilon ^4} \right.\nn
\\&\left.-\frac{4}{\epsilon ^3} +\frac{1}{\epsilon^2}\left(  \frac{2 \pi ^2}{3}-8   \right) +\frac{1}{\epsilon}\left( \frac{44 \zeta (3)}{3}-16   \right) +\frac{32 \zeta (3)}{3}+\frac{8 \pi ^4}{45}+\frac{2 \pi ^2}{3}-32  \right]\,,
\\
\mathcal{A}^{(1)a}_{( c- c)}=&\left(  \frac{\mu^2}{m_b^2}  \right)^{2\epsilon}\left( \frac{\nu^2}{m_b^2}   \right)^{\eta}\left[   \frac{1}{\eta}\left(  -\frac{2}{\epsilon ^3}-\frac{ \pi ^2}{ \epsilon } +\frac{16 \zeta (3)}{3} \right)+\frac{3}{\epsilon ^4}+\frac{2}{\epsilon ^3}+\frac{1}{\epsilon^2}\left( \frac{\pi ^2}{6} +4 \right) \right.\nn
\\&\left.   +\frac{1}{\epsilon}\left( \frac{\pi ^2}{3} +8 \right) -\frac{52 \zeta (3)}{3}-\frac{7 \pi ^4}{40}+\frac{2 \pi ^2}{3}+16     \right]\,,
\\
\mathcal{A}^{(1)a}_{( \bar c- \bar c)}=&\left(   \frac{\mu^2}{m_b^2}   \right)^{2\epsilon}\left(  -\frac{\nu^2}{m_H^2}     \right)^{\eta}\left[ \frac{1}{\eta}\left(  \frac{2}{\epsilon ^3}+\frac{ \pi ^2}{ \epsilon } -\frac{16 \zeta (3)}{3}   \right) +\frac{2}{\epsilon ^4}+\frac{2}{\epsilon ^3}+\frac{4}{\epsilon^2}   \right.\nn
\\&\left.    +\frac{1}{\epsilon}\left(-\frac{10 \zeta (3)}{3}+\frac{\pi ^2}{3}+8\right)-\frac{52 \zeta (3)}{3}-\frac{\pi ^4}{36}+\frac{2 \pi ^2}{3}+16    \right]\,.
\end{align} 
We find that the $1/\eta$ poles cancel between the $(c-h)$ and $(\bar c -h )$ regions, and between the $(c-c)$ and $(\bar c -\bar c)$ regions, respectively.

Adding up all the regions, we get
\begin{align}
\mathcal{A}^{(1)a}=&\mathcal{A}^{(1)a}_{(h-h)}+\mathcal{A}^{(1)a}_{( c- h)}+\mathcal{A}^{(1)a}_{( c- c)}+\mathcal{A}^{(1)a}_{( \bar c- h)}+\mathcal{A}^{(1)a}_{( \bar c- \bar c)}\nn
\\
=&  \left( -\frac{\mu^2}{m_H^2} \right)^{2\epsilon}  
\left[\frac{1}{\epsilon}\Big(    4 \ln ^2(-R)-16    \Big)   -\frac{1}{6}\ln^{4}(-R)-4\ln^{3}(-R) -\left(\frac{4\pi^{2}}{3}-6\right)\ln^{2}(-R) \right.\nn
\\&
\left.   -\left(32\zeta(3)+\frac{4\pi^{2}}{3}\right)\ln(-R)-56\zeta(3)-\frac{2\pi^{4}}{5}+\frac{10\pi^{2}}{3}-94  \right],
\end{align}
which agrees with the corresponding full result in eq. (\ref{eq:expand_NLOa}) at LP.
It is remarkable to see that the poles of $1/\epsilon^i, i=2,3,4$ are all canceled.

\subsubsection{Diagram \ref{fig:two_loop_dia}(b)}

\begin{figure}[t]
 \centering
 \vspace{0cm}
 \subfloat[]
{\includegraphics[width=0.3\textwidth]{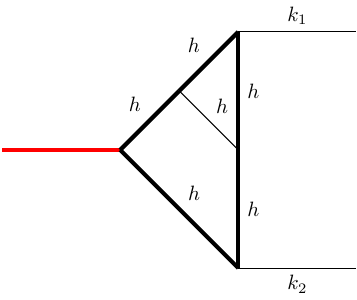}}
\subfloat[]
{\includegraphics[width=0.3\textwidth]{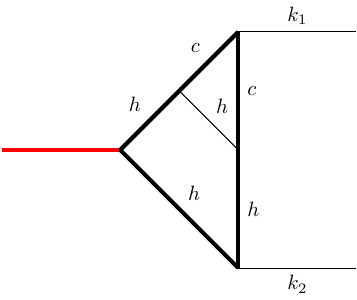}}
 \subfloat[]
{\includegraphics[width=0.3\textwidth]{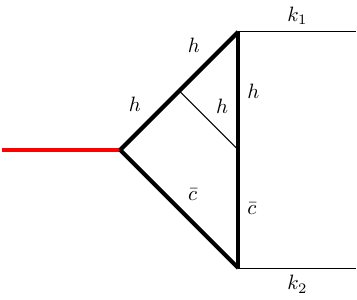}}

\subfloat[]
{\includegraphics[width=0.3\textwidth]{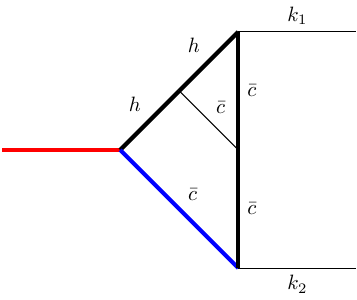}}
 \subfloat[ ]
{\includegraphics[width=0.3\textwidth]{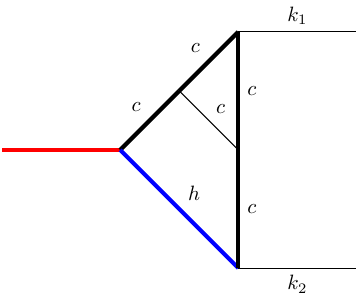}}
\caption { Five regions at LP for the diagram~\ref{fig:two_loop_dia}(b). 
The momentum scaling mode of each propagator is shown explicitly by the labels. An analytic regulator is imposed on the blue line.}
\label{fig:dia_b_eta}
\end{figure}

Then we consider the diagram \ref{fig:two_loop_dia}(b).
There are five LP regions contributing to the amplitude,
which are shown in figure \ref{fig:dia_b_eta}.
The integral in the hard region can be calculated directly, giving the result below,
\begin{align}
\mathcal{A}^{(1)b}_{(h-h)}=&  \left( -\frac{\mu^2}{m_H^2} \right)^{2\epsilon}  
\left[- \frac{2}{\epsilon^2}  - \frac{1}{\epsilon} \left(    \frac{2 \pi ^2}{3}  +8     \right) -4 \zeta (3)-2 \pi ^2-30        \right]\,.
\end{align}

The $(h-c)$ region gives a finite contribution, 
\begin{align}
\mathcal{A}^{(1)b}_{(h-c)}=-\frac{4\pi^2}{3} \,.
\end{align}
To understand this result, we perform the calculation of the hard one-loop amplitude firstly,
\begin{align}
\mathcal{M}^{(1)b,{\rm hard}}_{(h-c)}=ie_bg_s^2y_bN_cC_F
\int \frac{d^D l }{(2\pi)^D} \frac{\gamma^{\rho} \lsl \gamma^{\nu}_{\perp} (\lsl+\ksl_2)(\lsl-\ksl_1)\gamma_{\rho} }{l^2 (l+k_2)^2 (l-k_1)^2 (l-\tilde{q} )^2}
\end{align}
with $\tilde{q}^{\mu}=\bar{n}\cdot q n^{\mu}/2$.
We should also keep in mind that the above matrix element is sandwiched between two $\nsl$'s.
Therefore, the numerator can be simplified as
\begin{align}
  &  \gamma^{\rho}_{\perp} \lsl \gamma^{\nu}_{\perp} (\lsl+\ksl_2)(\lsl-\ksl_1)\gamma_{\perp\rho} \nn \\
  =& l^2 \frac{n\cdot l}{2}\gamma^{\rho}_{\perp} \gamma^{\nu}_{\perp} \gamma_{\perp\rho} \nbsl +
  \frac{m_H}{2}\gamma^{\rho}_{\perp} \lsl_{\perp} \gamma^{\nu}_{\perp}  \lsl_{\perp}\gamma_{\perp\rho}\nbsl \nn \\  
  =& \left( \epsilon l^2 (n\cdot l) +
  \frac{\epsilon^2}{1-\epsilon} m_H  l_{\perp}^2 \right) \gamma^{\nu}_{\perp}  \nbsl 
  \label{eq:examp}
\end{align}
where we have used $\gamma^{\rho}_{\perp}\gamma_{\perp \nu}\gamma_{\perp\rho}=2\epsilon \gamma_{\perp \nu}$ and $l_{\perp}^{\alpha} l_{\perp}^{\beta} = l_{\perp}^2 g_{\perp}^{\alpha\beta}/(2-2\epsilon)$ under integration.
Then the scalar integrals can be computed in Feynman parameter representation, and the full results in $\epsilon$ are given by
\begin{align}
    &\int \frac{d^D l }{(2\pi)^D} \frac{n\cdot l}{(l+k_2)^2 (l-k_1)^2 (l-\tilde{q} )^2}  \nn
    \\&= \frac{i(4\pi)^{\epsilon-2}(-m_H^2)^{-\epsilon}\Gamma(1-\epsilon)^2\Gamma(\epsilon)}{\Gamma(2-2\epsilon)(m_H-\bar n \cdot q)}\left[1-\left(  \frac{m_H}{\bar n\cdot q}  \right)^{\epsilon}  \right] , \\
    &\int \frac{d^D l }{(2\pi)^D} \frac{l_{\perp}^2 }{l^2 (l+k_2)^2 (l-k_1)^2 (l-\tilde{q} )^2} \nn
    \\&= \frac{im_H(4\pi)^{\epsilon-2}(-m_H^2)^{-\epsilon-1}\Gamma(2-\epsilon)\Gamma(-\epsilon)\Gamma(\epsilon)}{\Gamma(2-2\epsilon)(m_H-\bar n\cdot q)  }\left[1-\left(  \frac{m_H}{\bar n\cdot q}  \right)^{\epsilon}  \right] .   
\end{align}
We firstly note that the poles in $\Gamma(\epsilon)$ and $\Gamma(-\epsilon)\Gamma(\epsilon)$ cancel their corresponding coefficients $\epsilon$ and $\epsilon^2$ in  eq. (\ref{eq:examp}), respectively.
Then there is no longer an endpoint singularity at $\bar{n}\cdot q=0$.
Consequently, the remaining collinear $q$ integral gives rise to only an ultraviolet divergence $1/\epsilon$,
which would cancel against the terms in the above brackets once expanded in $\epsilon$.
Thus only a finite term is obtained in this region.
We also note that the denominator $(m_H-\bar{n}\cdot q)$ does not induce a new endpoint divergence due to the vanishing factor in the bracket in the limit of $\bar{n}\cdot q = m_H$.

In the $(\bar c-h)$ region,  we do not need an endpoint regulator either because the hard loop amplitude changes the endpoint singularity so that it is already regulated by the dimensional regulator.
At the end we obtain
\begin{align}
\mathcal{A}^{(1)b}_{(\bar c-h)}=&  \left( -\frac{\mu^2}{m_H^2} \right)^{\epsilon} \left( \frac{\mu^2}{m_b^2} \right)^{\epsilon}  
\left[\frac{2}{\epsilon ^3}+\frac{8}{\epsilon ^2}+\frac{16}{\epsilon }-\frac{16 \zeta (3)}{3}+32 \right]\,,
\end{align}
where $m_H$ and $m_b$ indicate the hard and anti-collinear scales, respectively.

The integrals in  the $(\bar c-\bar c)$ and $( c- c)$ regions suffer from endpoint divergences that cannot be regulated by the dimensional regulator.
As in the LO, we improve the power of one propagator from 1 to $1+\eta$, and calculate the loop integrals to obtain
\begin{align}
\mathcal{A}^{(1)b}_{(\bar c-\bar c)}=&  \left( \frac{\mu^2}{m_b^2} \right)^{2\epsilon}    \left( \frac{\nu^2}{m_b^2} \right)^{\eta}
\left[  \frac{1}{\eta}\left(   \frac{2}{\epsilon ^2}+\frac{4}{\epsilon }-\frac{\pi ^2}{3}+8  \right)   -\frac{1}{\epsilon^3}-\frac{2}{\epsilon ^2} +\frac{1}{\epsilon}\left(  \frac{\pi ^2}{6}-4  \right) + \frac{20 \zeta (3)}{3}\right.\nn
\\&\left.   +\frac{\pi ^2}{3}-8     \right]\,,  \\
\mathcal{A}^{(1)b}_{( c- c)}=&  \left( \frac{\mu^2}{m_b^2} \right)^{2\epsilon}    \left( -\frac{\nu^2}{m_H^2} \right)^{\eta}
\left[  \frac{1}{\eta}\left(   -\frac{2}{\epsilon ^2}-\frac{4}{\epsilon }+\frac{\pi ^2}{3}-8 \right)  -\frac{1}{\epsilon ^3}-\frac {4} {\epsilon^2}   +\frac{1}{\epsilon}\left(  \frac{\pi ^2}{2}-8   \right)    \right.\nn
\\&\left.   -\frac{10 \zeta (3)}{3}+\pi ^2-16  \right] \,.
\end{align}
The $\eta$ poles cancel between these two regions.

Adding up all the regions, we get
\begin{align}
\mathcal{A}^{(1)b}=&  \left( -\frac{\mu^2}{m_H^2} \right)^{2\epsilon}  
\left[\frac{1}{\epsilon} \Big(      \ln^2 (-R) - 4    \Big)  -\frac {5} {3} \ln^3 (-R) - \pi^2\ln (-R) - 6\zeta (3) - 2\pi^2 - 22    \right]
\end{align}
which agrees with the full result in eq. (\ref{eq:expand_NLOb}) at LP.
All the $1/\epsilon^i, i=2,3,$ poles cancel out.

\subsubsection{Diagram \ref{fig:two_loop_dia}(d)}

\begin{figure}[H]
 \centering
 \vspace{0cm}
 \subfloat[]
{\includegraphics[width=0.3\textwidth]{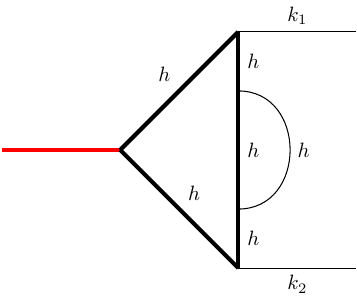}}
\subfloat[]
{\includegraphics[width=0.3\textwidth]{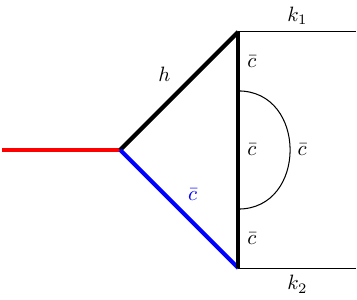}}
 \subfloat[]
{\includegraphics[width=0.3\textwidth]{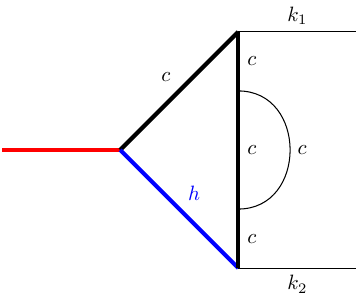}}
\caption { Three regions at LP for the diagram~\ref{fig:two_loop_dia}(d). 
The momentum scaling mode of each propagator is shown explicitly by the labels. An analytic regulator is imposed on the blue line.}
\label{fig:dia_d_eta}
\end{figure}

The diagram \ref{fig:two_loop_dia}(d) has only three LP regions, as shown in figure \ref{fig:dia_d_eta}.

As before, the hard region can be calculated directly, and we obtain
\begin{align}
\mathcal{A}^{(1)d}_{(h-h)}=&  \left( -\frac{\mu^2}{m_H^2} \right)^{2\epsilon}  
\left[ \frac{1}{\epsilon^3}  + \frac{2}{\epsilon^2} + \frac{1}{\epsilon}\left(  \frac{\pi ^2}{6}  +7   \right)  -\frac{26 \zeta (3)}{3}+\frac{\pi ^2}{3}+\frac{49}{2}       \right]\,.
\end{align}
The other two regions require the introduction of an endpoint regulator.
We choose the same method as the LO to impose the regulator.
After calculation of the relevant integrals in Feynman parameters,
we obtain the results
\begin{align}
\mathcal{A}^{(1)d}_{(\bar c-\bar c)}=&  \left( \frac{\mu^2}{m_b^2} \right)^{2\epsilon}    \left( \frac{\nu^2}{m_b^2} \right)^{\eta}
\left[  \frac{1}{\eta}\left(  \frac{2}{\epsilon ^2}-\frac{8}{\epsilon }+\pi ^2-8    \right)  -\frac{1}{\epsilon^3}  -\frac{5}{\epsilon^2} +\frac{1}{\epsilon} \left(   \frac{\pi ^2}{6}  +6      \right)   -\frac{28 \zeta (3)}{3}\right.\nn
\\&\left.   -\frac{11 \pi ^2}{6}+4            \right]\,,\\
\mathcal{A}^{(1)d}_{(c-c)}=&  \left( \frac{\mu^2}{m_b^2} \right)^{2\epsilon}    \left( -\frac{\nu^2}{m_H^2} \right)^{\eta}
\left[  \frac{1}{\eta}\left(  -\frac{2}{\epsilon ^2}+\frac{8}{\epsilon }-\pi ^2+8   \right)   -  \frac{3}{\epsilon^2}   +   \frac{1}{\epsilon}\left( -\frac{\pi ^2}{3}  +10  \right)   -2 \zeta (3)\right.\nn
\\&\left.   -\frac{5 \pi ^2}{6}+12    \right]\,.
\end{align}
It can be seen that the $1/\eta$ poles cancel exactly.
The sum of all three regions is given by
\begin{align}
\mathcal{A}^{(1)d}=&  \left( -\frac{\mu^2}{m_H^2} \right)^{2\epsilon}  
\left[- \frac{6}{\epsilon^2} + \frac{1}{\epsilon}\Big(   2\ln^2 (-R) + 24\ln (-R) + 23      \Big)       -\frac {8} {3} \ln^3 (-R) 
\right.\nn 
\\&\left.    
-  32\ln^2 (-R) - \left (  \frac {2\pi^2} {3}+24 \right)\ln (-R) - 
 20\zeta (3)  - \frac {7\pi^2} {3} + \frac {81} {2}   \right]\,,
\end{align}
which agrees with the full result in eq. (\ref{eq:expand_NLOd}) at LP.
We also find that the $1/\epsilon^3$ cancel in the sum.

\subsubsection{Diagram \ref{fig:two_loop_dia}(e)}

\begin{figure}[t]
 \centering
 \vspace{0cm}
 \subfloat[]
{\includegraphics[width=0.3\textwidth]{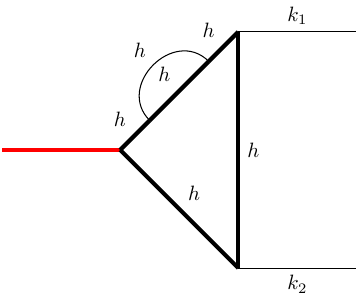}}
\subfloat[]
{\includegraphics[width=0.3\textwidth]{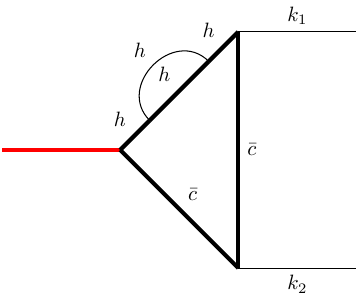}}
 \subfloat[]
{\includegraphics[width=0.3\textwidth]{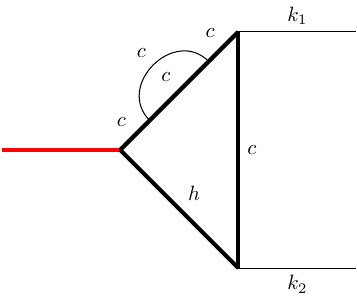}}
\caption {Three regions at LP for the diagram~\ref{fig:two_loop_dia}(e). 
The momentum scaling mode of each propagator is shown explicitly by the labels.}
\label{fig:dia_e_eta}
\end{figure}

The diagram \ref{fig:two_loop_dia}(e) is similar to the diagram \ref{fig:two_loop_dia}(d)
since they both contain a self-energy loop correction.
The three regions giving LP contributions are shown in figure \ref{fig:dia_e_eta}.
However, they have different analytical structures.
Specifically, the self-energy loop correction in figure \ref{fig:dia_d_eta}(c) does not change the endpoint singularity and an additional regulator except for the dimensional regulator is needed.
In contrast, the self-energy loop correction in figure \ref{fig:dia_e_eta}(c) generates $(\bar n\cdot l)^{\epsilon}$ in the limit $\bar n\cdot l \to 0$ after integration over $n\cdot l$ using the residue theorem, which regulates the endpoint singularity.
The self-energy loop correction in figure \ref{fig:dia_e_eta}(b) is proportional to  $(n\cdot l)^{-\epsilon }$, making the integration over $(n\cdot l)$ well-defined already in dimensional regulator.
Therefore it is not necessary to introduce the $\eta$ regulator for this diagram.

The analytical results for the three regions are given by
\begin{align}
\mathcal{A}^{(1)e}_{(h-h)}=&  \left( -\frac{\mu^2}{m_H^2} \right)^{2\epsilon}  
\left[   -  \frac{1}{\epsilon^3} -\frac{1}{\epsilon^2}+\frac{1}{\epsilon}\left(  \frac{\pi ^2}{6}-\frac{7}{2}  \right)  +  \frac{32 \zeta (3)}{3}+\frac{\pi ^2}{6}-\frac{61}{4}    \right]\,,
\\
\mathcal{A}^{(1)e}_{(\bar c-h)}=&  \left( -\frac{\mu^2}{m_H^2} \right)^{\epsilon}\left( \frac{\mu^2}{m_b^2} \right)^{\epsilon}    
\left[     \frac{2}{\epsilon^3} +\frac{2}{\epsilon^2}+\frac{4}{\epsilon}   -\frac{16 \zeta (3)}{3}  + 8  \right] \,, \\
\mathcal{A}^{(1)e}_{(c-c)}=&  \left( \frac{\mu^2}{m_b^2} \right)^{2\epsilon}  
\left[     -\frac{1}{\epsilon^3} +\frac{2}{\epsilon^2} -\frac{1}{\epsilon}\left(  12+\frac{\pi ^2}{6}     \right)          +  \frac{74 \zeta (3)}{3}+\frac{\pi ^2}{3}-16   \right]\,.
\end{align}
Their sum reads
\begin{align}
\mathcal{A}^{(1)e}=&\mathcal{A}^{(1)e}_{(c-c)}+\mathcal{A}^{(1)e}_{(\bar c-h)}+\mathcal{A}^{(1)e}_{(h-h)}
 \\=& \left( -\frac{\mu^2}{m_H^2} \right)^{2\epsilon}  
\left[  \frac{3}{\epsilon^2}-\frac{1}{\epsilon}\left(   \ln^2 (-R) + 6\ln (-R) + \frac {23} {2}    \right)  +   \ln^3 (-R) + 
 5\ln^2 (-R)  \right.\nn
 \\&\left.  + \left ( \frac {\pi^2} {3}+20 \right)\ln (-R) +30 \zeta (3)+\frac{\pi ^2}{2}-\frac{93}{4}  \right]\,,
\end{align}
which agrees with the full result in eq. (\ref{eq:expand_NLOe}) at LP. Again, we find that the $1/\epsilon^3$ poles cancel out as expected.

\section{Calculation with $\Delta$ regulators}
\label{sec:Delta}

In the last section, we have used the analytic regulator for the endpoint singularity.
The relevant scales in each region are clearly recognized from the corresponding result.
And summing all the regions can reproduce the correct result in the full theory for each diagram.
However, the LO results in this regulator are divergent in different regions.
And the ultraviolet divergences are entangled with the endpoint divergences; see the discussion around eq. (\ref{eq:AcbarLO}).
In order to have finite LO results and to isolate the divergences of different origins, it is desirable to employ another kind of regulator in calculation.
In this section, we consider the $\Delta$-regulator \cite{Chiu:2009yx}.
Specifically, we modify the propagators that could develop an endpoint singularity after power expansion, 
for example,
\beq
\begin{aligned}
\frac{1}{(l-k_1)^2}\to \frac{1}{(l-k_1)^2+\Delta_1}\,,\quad
\frac{1}{(l+k_2)^2}\to \frac{1}{(l+k_2)^2+\Delta_2}\,.
\end{aligned}
\eeq
Note that the two $\Delta$'s do not have an explicit scaling in the power counting parameter $\lambda$.
Therefore they are not dropped after power expansion.
With such a regulator, the endpoint singularity manifests itself in the form of $\ln \delta_i $ with $\delta_i=\Delta_i/m_H^2$, while
the ultraviolet divergences are still in the form of $1/\epsilon^i$.
Another feature of this regulator is that the soft regions are not scaleless so that we can investigate all the regions in depth.

\subsection{One-loop amplitudes}

\begin{figure}[H]
 \centering
 \vspace{0cm}
 \subfloat[]
{\includegraphics[width=0.23\textwidth]{oneloop_region3.pdf}}
\subfloat[]
{\includegraphics[width=0.23\textwidth]{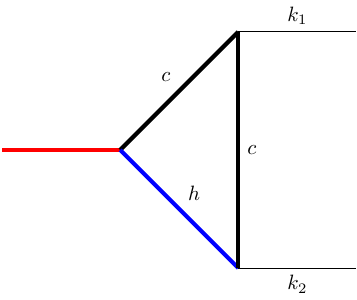}}
 \subfloat[]
{\includegraphics[width=0.23\textwidth]{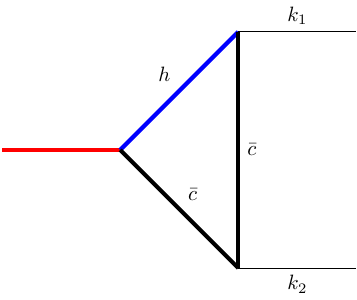}}
 \subfloat[]
{\includegraphics[width=0.23\textwidth]{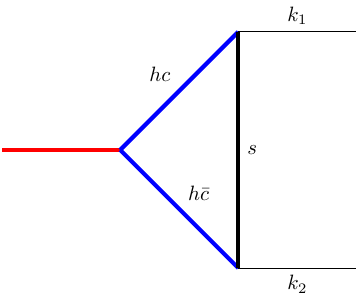}}
\caption {Four regions in the one-loop contribution when we use the $\Delta$ regulator. 
The momentum scaling mode of each propagator is shown explicitly by the labels ($h$,  $hc$, $h\bar{c}$, $c$, $\bar{c}$,  and $s$). The propagator with a $\Delta$ is in blue.}
\label{fig:oneloop_region_delta}
\end{figure}

At LO, the one-loop integral receives contributions from  four regions, which are shown in figure ~\ref{fig:oneloop_region_delta}.
The integral in the hard region does not suffer from the endpoint divergence and thus the result remains the same as in eq. (\ref{eq:one_loop_hard_region_eta}).

In the collinear region, the endpoint singularity appears at $\bar{n}\cdot l =0$.
Therefore, we keep $\Delta_2$ in the corresponding propagator
and drop the other $\Delta$'s.
The amplitude in this region is given by
\begin{align}
    \mathcal{A}_{c}^{(0)}=& -32i\pi^2m_H^2  \int\frac{d^Dl}{(2\pi)^D}\frac{(\tilde{\mu}^2)^\epsilon}{(l^2-m_b^2+i0)[(l-k_1)^2-m_b^2+i0][m_H(\bar n\cdot l)+\Delta_2+i0]}\nn
    \\&= \left(  \frac{\mu^2}{m_b^2}     \right)^{\epsilon}\bigg[-\frac{2}{\epsilon}\ln\delta_2 +\mathcal{O}(\epsilon) +\mathcal{O}(\delta_2) \bigg]\,.
\end{align}
Comparing it with the result in eq. (\ref{eq:AcLO}), we see that the endpoint singularity changes from $1/\eta$ to $\ln \delta_2  $ which involves only hard scales.
Since we will take the limit $\Delta_2 \to 0$ at the end, 
we neglect the $\mathcal{O}(\delta_2)$ terms in the results.

Similarly, we obtain the result in the anti-collinear region,
\begin{align}
    \mathcal{A}_{\bar c}^{(0)}=& -32i\pi^2m_H^2  \int\frac{d^Dl}{(2\pi)^D}\frac{(\tilde{\mu}^2)^\epsilon}{(l^2-m_b^2+i0)[-m_H(n\cdot l)+\Delta_1+i0][(l+k_2)^2-m_b^2+i0]}\nn
    \\=& \left(  \frac{\mu^2}{m_b^2}     \right)^{\epsilon}\bigg[ -\frac{2}{\epsilon}\ln\delta_1 +\mathcal{O}(\epsilon) +\mathcal{O}(\delta_1) \bigg]\,,
\end{align}
which is the same as the collinear result after replacing $\delta_2$ by $\delta_1$.
This symmetry holds even at higher orders
so that one does not need to calculate the anti-collinear result. 
This is in contrast with the situation in the analytic regulator discussed in the last section\footnote{In other analytic regulators, the collinear and anti-collinear regions may give the same contribution; see ref.~\cite{Liu:2019oav}.}
where the collinear and anti-collinear results have to be calculated independently.
The mismatch between the two kinds of regulators indicates that there must be additional regions in $\Delta$ regulators.

Indeed, the soft region, in which the loop momentum is scaling as $l^{\mu}\sim (\lambda, \lambda ,\lambda)m_H$, gives a non-vanishing result,
\begin{align}
    \mathcal{A}_{s}^{(0)}=& -32i\pi^2m_H^2  \int\frac{d^Dl}{(2\pi)^D}\frac{(\tilde{\mu}^2)^\epsilon}{(l^2-m_b^2+i0)[-m_H(n\cdot l)+\Delta_1+i0][m_H(\bar n\cdot l)+\Delta_2+i0]} \nn
    \\=&\left(  \frac{\mu^2}{m_b^2}     \right)^{\epsilon}\Bigg[\frac{2}{\epsilon^2}-\frac{2}{\epsilon}\Big( \ln \frac{\Delta_1}{m_H m_b}+\ln \frac{\Delta_2}{m_H m_b} -i\pi  \Big)  -\frac{\pi^2}{6} +\mathcal{O}(\epsilon)+\mathcal{O}(\Delta_1)+\mathcal{O}(\Delta_2)  \Bigg]\,.
\end{align}
The prefactor  $(\mu^2/m_b^2)^\epsilon$ indicates the scale of the soft loop momentum.
The endpoint logarithms in the bracket involve the scale of $\mathcal{O}(\lambda)$ because the endpoint integral is carried out for the light-cone components $n\cdot l$ or $\bar{n}\cdot l$,
which are of order $\lambda$.
Comparing the above three equations, we find that $\Delta_i, i=1,2,$ have different scaling behaviors in different regions.
This is not a problem in the application of the method of regions
because $\Delta$'s serve as regulators rather than intrinsic scales in the integral.

We stress that the statement that 
the successive expansions in two different regions give a scaleless and  vanishing integral does not hold any more with $\Delta$ regulators. 
The expansion of the integrand of 
$ \mathcal{A}_{c}^{(0)} $ in the soft (or anti-collinear) region 
yields just the same integrand of  $\mathcal{A}_{s}^{(0)}$.
We have to subtract this contribution to avoid double counting \cite{Manohar:2006nz,Idilbi:2007ff,Idilbi:2007yi}, and derive the result for the one-loop amplitude,
\begin{align}
    \mathcal{A}^{(0)}=&\mathcal{A}^{(0)}_{s}+(\mathcal{A}^{(0)}_{c}-\mathcal{A}^{(0)}_{s})+(\mathcal{A}^{(0)}_{\bar c}-\mathcal{A}^{(0)}_{s})+\mathcal{A}^{(0)}_{h} \nn
    \\=& \mathcal{A}^{(0)}_{c}+\mathcal{A}^{(0)}_{\bar c}-\mathcal{A}^{(0)}_{s}+\mathcal{A}^{(0)}_{h}  \nn
    \\=&   \ln^2 R+2i\pi \ln R -\pi^2 -4\,,
\end{align}
where all divergences in $1/\epsilon^i$ and $\ln \delta_i$ cancel out.
We find complete agreement with eq. (\ref{eq:Aexp}) obtained in an analytic regulator.


In the above method, we have separated the ultraviolet and endpoint divergences in the collinear and soft regions.
However, the LO result in each region is still divergent in $\epsilon$.
This divergence is related to the ultraviolet divergence in the transverse loop momentum integration.
In order to isolate such a divergence, we split the integration region of the transverse momentum into two parts, i.e., 
$0\le l_T \le m_H$ and $l_T>m_H$, where we have used  $l_{\perp}^2=-l_T^2$.
Consequently, we write the amplitude as
\beq
\begin{aligned}
\mathcal{A}=&\mathcal{A}_{c,l_T\le m_H}+\mathcal{A}_{\bar c, l_T \le m_H}-\mathcal{A}_{s,l_T\le m_H}
\\ +&\mathcal{A}_{c, l_T>m_H}+\mathcal{A}_{\bar c, l_T>m_H}-\mathcal{A}_{s, l_T>m_H}+\mathcal{A}_h\,.
\label{eq:apart_A}
\end{aligned}
\eeq
Now the terms in the first line are finite at LO.
All the ultraviolet divergences in the collinear and soft regions appear merely in the second line and would cancel against the infrared divergences in the hard region.
The requirement of $l_T > m_H $ ensures that the small scale $m_b$ can be dropped in the second line, and thus only hard scale is involved.

Explicit calculations lead to the following LO results:
\begin{align}
\mathcal{A}^{(0)}_{c,l_T\le m_H} & =2(\ln R )\ln  \delta_2\,,\label{eq:Ac0small}\\
\mathcal{A}^{(0)}_{\bar c,l_T\le m_H} &=2 (\ln R) \ln  \delta_1\,, \\
\mathcal{A}^{(0)}_{s,l_T\le m_H} & =  2 (\ln R) \ln \delta_1+2 (\ln R) \ln \delta_2  -\ln^2 R-2i\pi\ln R  \,,
\label{eq:As0small}
\end{align}
and 
\begin{align}
    \mathcal{A}^{(0)}_{\bar c,l_T> m_H} & =\left(  \frac{\mu^2}{m_H^2} \right)^{\epsilon}\bigg[ -\frac{2}{\epsilon}\ln \delta_1  +\mathcal{O}(\epsilon)+\mathcal{O}(\delta_1)\bigg] \,, \\
    \mathcal{A}^{(0)}_{c,l_T> m_H} & = \left(  \frac{\mu^2}{m_H^2} \right)^{\epsilon}\bigg[ -\frac{2}{\epsilon}\ln \delta_2  +\mathcal{O}(\epsilon)+\mathcal{O}(\delta_2)\bigg] \,, \\
    \mathcal{A}^{(0)}_{s,l_T> m_H} & =\left(  \frac{\mu^2}{m_H^2} \right)^{\epsilon}\bigg[ \frac{2}{\epsilon^2}-\frac{2}{\epsilon}\bigg( \ln \delta_1 +\ln \delta_2 -i\pi  \bigg)  -\frac{\pi^2}{6} +\mathcal{O}(\epsilon)+\mathcal{O}(\delta_i)  \bigg] \,.
\end{align}
The LO results are indeed finite in the parts with $l_T\le m_H$,
and the symmetry between the collinear and anti-collinear regions holds.
The sum of the large $l_T$ parts in the above equations and $\mathcal{A}_h^{(0)}$ contains no scales, contributing only constants. 
All the logarithms are encompassed in the small $l_T$ parts.
Summarising eqs. (\ref{eq:Ac0small}-\ref{eq:As0small}) gives 
\beq
\begin{aligned}
\mathcal{A}^{(0)}_{c,l_T\le m_H}+\mathcal{A}^{(0)}_{\bar c,l_T\le m_H}-\mathcal{A}^{(0)}_{s,l_T\le m_H}
=    \ln^2 R+2i\pi \ln R  \,,
\end{aligned}
\eeq
which agrees with eq. (\ref{eq:LOA}).
Note that the logarithms can be considered only from  the soft region if we choose $\delta_i=1$.

\subsection{Two-loop amplitudes}

In this section, we will demonstrate that all the logarithms can be reproduced by calculating the two-loop amplitudes only in the small $l_T$ region.
Before we proceed to calculate the relevant integrals,
we should firstly determine the regions for each diagram with the $\Delta$ regulators.
As discussed in the last subsection, the parameter $\Delta_i$ does not have a definite power counting 
and therefore it is not appropriate to search for all regions with this regulator in the integral. 
Instead, we add an auxiliary propagator $1/(k_1\cdot l+k_2\cdot l)^{\eta}$ in the integrand and find all the regions by using {\tt Asy2.1} \cite{Jantzen:2012mw} and {\tt pySecDec} \cite{Heinrich:2021dbf}.
The momentum modes of the propagators are then derived following the method described at the beginning of section \ref{sec:expansion}.
Note that we have chosen such an auxiliary propagator on purpose. 
For the regions in which the momentum $l$ is considered hard,  or (anti-)collinear, the leading power of the denominator $k_1\cdot l + k_2\cdot l $ scales as $\mathcal{O}(m_H^2)$.
Taking the limit $\eta\to 0$ reproduces the same hard sub-graph and thus the same regions as those in the analytic regulator discussed above.
For the other regions, the momentum $l$ is soft.
The original integral (the integral with $\eta=0$) is scaleless but now brought out using the auxiliary propagator.
The very correspondence between the $l$-soft  and $l$-collinear regions, i.e.,  the integral in the soft region can be obtained after expansion of the collinear integral in the soft or anti-collinear region, guarantees that all the relevant regions have been identified.
Then we apply the $\Delta$ regulators and
the integrals in each region are calculated loop-by-loop due to the constraint on $l_T$.

\subsubsection{Diagram \ref{fig:two_loop_dia}(a)}

\begin{figure}[ht]
 \centering
 \vspace{-0.3cm}
  \subfloat[]
{\includegraphics[width=0.25\textwidth]{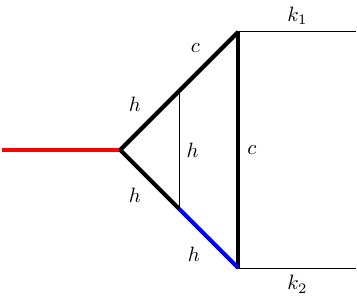}}
\subfloat[]
{\includegraphics[width=0.25\textwidth]{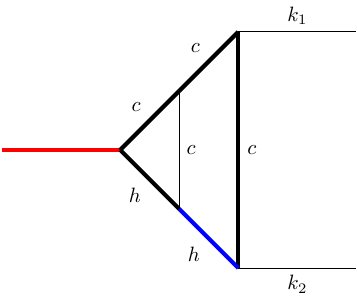}}
 \subfloat[]
{\includegraphics[width=0.25\textwidth]{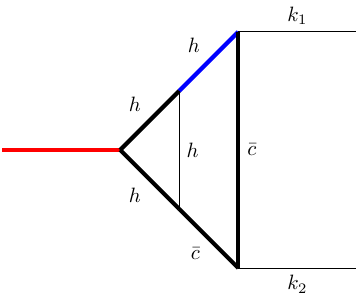}}
\subfloat[]
{\includegraphics[width=0.25\textwidth]{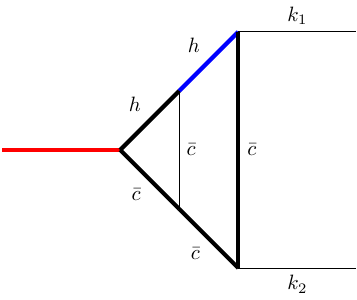}}\\
 \subfloat[]
{\includegraphics[width=0.25\textwidth]{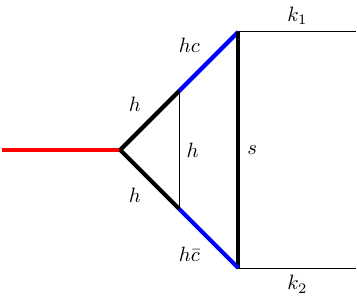}}
\subfloat[]
{\includegraphics[width=0.25\textwidth]{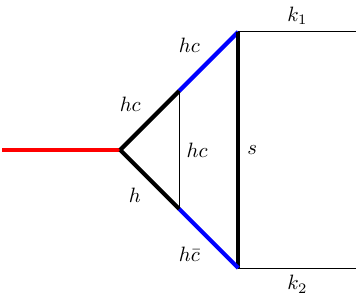}}
 \subfloat[]
{\includegraphics[width=0.25\textwidth]{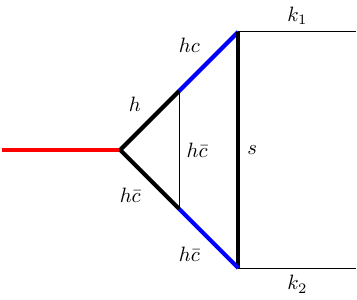}}
\subfloat[]
{\includegraphics[width=0.25\textwidth]{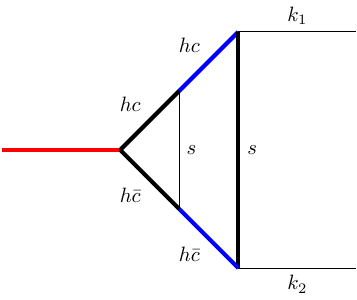}}
\caption{Regions contributing to LP logarithms for the diagram \ref{fig:two_loop_dia}(a) when we use the $\Delta$ regulator.
The propagator with a $\Delta$ is in blue.}
\label{fig:dia_a_region_delta}
\end{figure}

The regions of the diagram \ref{fig:two_loop_dia}(a)  that can contribute to LP logarithms are shown in figure \ref{fig:dia_a_region_delta}.

Here we take the $(c-h)$ region as an example.
The main integral appearing in this region is 
similar to eq. (\ref{eq:two_loop_example}) but with the $\eta$ regulator replaced by the $\Delta$ regulator,
\begin{align}
I_{(c-h),l_T\le m_H}&=\int_{l_T\le m_H} \frac{d^Dl}{(2\pi)^D} \frac{d^Dq}{(2\pi)^D} \frac{1}{(l^2-m_b^2+i0)[l^2-m_H(n\cdot l)-m_b^2+i0]} \nn
\\
&\times \frac{1}{[m_H(\bar n\cdot l)+\Delta_2+i0](q^2+i0)[q^2-m_H(n\cdot q)+(\bar n\cdot l)(n\cdot q)+i0]}\nn
\\
&\times \frac{1}{[q^2+m_H(\bar n\cdot q)+(\bar n\cdot l)(n\cdot q)+m_H(\bar n\cdot l)+i0]}.
\end{align}
We firstly integrate the hard loop momentum $q$ using Feynman parameters,
\begin{align}
&\int \frac{d^Dq}{(2\pi)^D} \frac{1}{(q^2+i0)[q^2-m_H(n\cdot q)+(\bar n\cdot l)(n\cdot q)+i0]}\nn
\\& \times \frac{1}{[q^2+m_H(\bar n\cdot q)+(\bar n\cdot l)(n\cdot q)+m_H(\bar n\cdot l)+i0]}\nn
\\=&\frac{i\big(-m_H^2  \big)^{-\epsilon}(e^{-\gamma_E}4\pi)^{\epsilon} }{16\pi^2m_H(m_H-\bar n\cdot l)} \bigg[ \bigg( \frac{m_H}{\bar n \cdot l}\bigg)^{\epsilon}  -1 \bigg]\left(-\frac{1}{\epsilon^2}+\frac{\pi^2}{12} +\mathcal{O}(\epsilon) \right)\,.
\label{eq:qhard-res}
\end{align}
Then we perform the integration over  $(n\cdot l)$ using the residue theorem, and obtain 
\begin{align}
    I_{(c-h),l_T\le m_H}&=\frac{(-m_H^2)^{-1-\epsilon}(e^{-\gamma_E}4\pi)^{\epsilon}}{(4\pi)^{4-\epsilon}\Gamma(1-\epsilon)}\int_{0}^{m_H^2} dl_{T}^2 l_T^{-2\epsilon}\int_0^{m_H} d(\bar n\cdot l)\frac{1}{(l_T^2+m_b^2)(m_H^2+\Delta_2)}      \nn
    \\& \times \Bigg[ \frac{m_H}{m_H(\bar n\cdot l)+\Delta_2} +  \frac{1}{m_H-\bar n\cdot l}     \Bigg]\bigg[  \left( \frac{m_H}{\bar n\cdot l}\right)^{\epsilon}-1   \bigg]
     \left(-\frac{1}{\epsilon^2}+\frac{\pi^2}{12} +\mathcal{O}(\epsilon) \right) \,.
    \label{eq:example_lTcut}
\end{align}
We have applied partial fraction in the bracket.
Notice that the second term does not generate a new endpoint divergence around $\bar{n}\cdot l= m_H$ due to the $[(m_H/\bar{n}\cdot l )^{\epsilon}-1]$ combination.
Because the endpoint divergence is regulated by $\Delta_2$, we can expand $(\bar{n}\cdot l)^{-\epsilon}$ in a series of $\epsilon$.
This is an advantage of using this regulator compared to the analytic regulator.
And we can also expand $l_T^{-2\epsilon}$ since the upper bound is finite and the lower bound does not cause any divergence.
Then making use of the following integration formula for the polylogarithm, 
\begin{align}
    \int_0^1 dt \frac{\ln^s t}{t+\delta} = (-1)^{s+1} s!~\text{Li}_{s+1}(-\delta^{-1}) , \quad s\in \mathbb{N}\,,
\end{align}
we perform the remaining integration over $\bar{n}\cdot l $ and $l_T^2$,
\begin{align}
    I_{(c-h),l_T\le m_H}
    &=\frac{(-m_H^2)^{-\epsilon}m_H^{-2\epsilon}(e^{-\gamma_E}4\pi)^{2\epsilon}}{256\pi^4m_H^4\epsilon}\Bigg[\text{Li}_1\left( -\frac{1}{R}\right) +\epsilon\text{Li}_2\left( -\frac{1}{R}\right)  \Bigg]        \nn
    \\& \times  \Bigg[ \text{Li}_2\left( -\frac{1}{\delta_2} \right)-\text{Li}_2(1)   +  \epsilon\bigg(  \text{Li}_3\left( -\frac{1}{\delta_2} \right)  - \text{Li}_3(1) \bigg) \Bigg]+\mathcal{O}(\epsilon)               \nn
    \\
   &=\frac{(-m_H^2)^{-\epsilon}m_H^{-2\epsilon}(e^{-\gamma_E}4\pi)^{2\epsilon}}{512\pi^4m_H^4} \Bigg[   -\frac{1}{\epsilon}\ln R\bigg( \ln^2 \delta_2+\frac{2\pi^2}{3}  \bigg)  \nn
    \\&+\ln^2R\bigg( \frac{1}{2}\ln^2\delta_2+\frac{\pi^2}{3}  \bigg) +\ln R\bigg(  \frac{1}{3}\ln^3\delta_2 +\frac{1}{3}\pi^2\ln\delta_2 -2\zeta(3) \bigg) \nn
    \\&+ \frac{1}{6}\pi^2\ln^2\delta_2 +\frac{\pi^4}{9} +\mathcal{O}(\epsilon)+\mathcal{O}(\delta_2)+\mathcal{O}(R)  \Bigg]\,.
\end{align}
In the last equation, we have expanded the polylogarithms in the limit of $\delta_2\to 0$ and $R\to 0$ using the following identity
\begin{align}
    \text{Li}_s(z) + (-1)^s \text{Li}_s(1/z) = - \frac{(2\pi i)^s}{s!}B_s \left(\frac{1}{2}+\frac{\ln(-z)}{2\pi i}\right)
\end{align}
with $B_s(x)$ being the Bernoulli polynomial.
We note that all the leading logarithms, i.e., the $\ln^i R \ln^{j}\delta_2$ terms with $i+j=4$, depend on $\delta_2$ while some of the next-to-next-to-leading logarithms i.e., the $\ln^i R \ln^{j}\delta_2$ terms with $i+j=2$, can be independent of $\delta_2$.
Consequently, the contribution from the leading logarithms is vanishing if we choose $\delta_2=1$.
After calculating the other integrals in the same way,
we obtain the amplitude in this region
\begin{align}
    \mathcal{A}^{(1)a}_{(c-h),l_T\le m_H}=&\left(     \frac{\mu^2}{m_H^2}   \right)^{2\epsilon} \Bigg[   \frac{1}{\epsilon}\Bigg(   \ln R \left(-2 \ln ^2\delta_2+4 \ln \delta_2-\frac{4 \pi ^2}{3}\right)    \Bigg) \nn
    \\&
    +\ln ^2R \left(\ln ^2\delta_2-2 \ln \delta_2+\frac{2 \pi ^2}{3}\right)+\ln R \left(\frac{2}{3} \ln ^3\delta_2-2 i \pi  \ln ^2\delta_2\right.\nn
    \\&\left.
    -2 \ln ^2\delta_2+\frac{2}{3} \pi ^2 \ln \delta_2+4 i \pi  \ln \delta_2+4 \ln \delta_2-4 \zeta (3)-\frac{4 i \pi ^3}{3}-\frac{2 \pi ^2}{3}\right)\nn
    \\&
      +\frac{1}{3} \pi ^2 \ln ^2\delta_2-\frac{2}{3} \pi ^2 \ln \delta_2 +\frac{2 }{9}\pi ^4   \Bigg]\,.
\end{align}

The amplitude in the $(c-c)$ region is more complicated.
Expansion of the numerator in this region leads to the terms that are independent of the loop momenta and the terms depending on the loop momenta in the form of $\bar{n}\cdot (l+q)$ .
The latter would cancel one of the denominators, making the integral simpler.
Below we show calculation techniques for the former, which are represented by
\begin{align}
I_{(c-c),l_T\le m_H}&=\int_{l_T\le m_H} \frac{d^Dl}{(2\pi)^D} \frac{d^Dq}{(2\pi)^D} \frac{1}{(l^2-m_b^2+i0)[l^2-m_H(n\cdot l)-m_b^2+i0]} \nn
\\
&\times \frac{1}{[m_H(\bar n\cdot l)+\Delta_2+i0](q^2+i0)[(l+q-k_1)^2-m_b^2+i0]}\nn
\\
&\times \frac{1}{[m_H\bar n \cdot (l+q)+i0]}.
\end{align}
We integrate $q$ using Feynman parameters firstly,
\begin{align}
&\int \frac{d^Dq}{(2\pi)^D} \frac{1}{(q^2+i0)[(l+q-k_1)^2-m_b^2+i0][m_H\bar n \cdot (l+q)+i0]}  \nn
\\=&\frac{-i(e^{-\gamma_{E}} 4\pi )^{\epsilon}m_H^{-2\epsilon}}{16\pi^2m_H(m_H-\bar n\cdot l)}\Bigg[ -\frac{1}{\epsilon}\ln\frac{m_H}{\bar n\cdot l} +  \text{Li}_2\left(   \frac{(\bar n\cdot l)\big(l^2-m_H(n\cdot l)\big)}{m_Hl^2-m_H^2(n\cdot l)+m_b^2(\bar n\cdot l-m_H)}     \right)   \nn
\\&-  \text{Li}_2\left(   \frac{m_H\big(l^2-m_H(n\cdot l)\big)}{m_Hl^2-m_H^2(n\cdot l)+m_b^2(\bar n\cdot l-m_H)}     \right) +\text{Li}_2\left(1-\frac{m_H}{\bar n\cdot l}   \right)  \nn
\\&+\ln   \frac{\bar n\cdot l }{m_H}     \ln \left( \frac{m_H^2( \bar n\cdot l-m_H)}{m_Hl^2-m_H^2(n\cdot l)+m_b^2(\bar n\cdot l-m_H)}   \right)  \Bigg]+\mathcal{O}(\epsilon)\,.
\label{eq:qcoll-res}
\end{align}
Note that there is no singularity at $\bar{n}\cdot l=m_H$.
We have neglected higher orders in $\epsilon$. They will not contribute to the final result because the integration of the loop momentum $l$ is finite in our setup. 
Then we perform the integration over  $(n\cdot l)$ using the residue theorem after analyzing the pole structure and branch cuts of the logarithms and polylogarithms, obtaining
\begin{align}
I_{(c-c),l_T\le m_H}&=\frac{2^{2\epsilon-8}\pi^{\epsilon-4}(e^{-\gamma_{E}} 4\pi )^{\epsilon}}{m_H^4\Gamma(1-\epsilon)}\int_{R}^{1+R} dx\int_0^{1} dz     \frac{m_H^{-4\epsilon}(x-R)^{-\epsilon}}{x(z+\delta_2)(1-z)}     \nn
\\& \times   \Bigg[ \frac{1}{\epsilon} \ln z      +   \text{Li}_2\left(   \frac{ z \big(  1 - Rz/x      \big)  }{1 - Rz^2 /x }   \right)  - \text{Li}_2\left(   \frac{  1 - R z /x     }{1 - Rz^2/x}   \right) +\text{Li}_2\left(1-\frac{1}{z}\right)  \nn
\\&  +\ln^2z+\ln z \ln(1-z)  -\ln z \ln (x-Rz^2)  \Bigg]\,.
\end{align}
For convenience we have substituted  $l_T^2+m_b^2=m_H^2 x$ and  $\bar n\cdot l=m_Hz$ in the calculation.
Because the integrand is a regular function around $z=1$, we expand the integrand in terms of $Rz/x$ or $Rz^2/x$ which can be considered always less than 1.
We only need to keep the leading terms in this expansion since higher-order terms cancel the endpoint singularity at $z=0$, causing no logarithms from the $z$ integration, and the integration over $x$ generates only constant contributions at $\mathcal{O}(\epsilon^0)$. 
After expansion, the integral $I_{(c-c),l_T\le m_H}$ can be calculated easily,
\begin{align}
I_{(c-c),l_T\le m_H}
&=\frac{2^{2\epsilon-8}\pi^{\epsilon-4}(e^{-\gamma_{E}} 4\pi )^{\epsilon}(m_b)^{-4\epsilon}}{m_H^4\Gamma(1-\epsilon)}      \Bigg[\frac{\ln R }{\epsilon} \bigg(  \frac{1}{2}\ln^2 \delta_2   +\frac{1}{3}\pi^2    \bigg) \nn
\\&  +\ln^2R\left(  \frac{1}{2}\ln^2\delta_2+\frac{\pi^2}{3}  \right) +\ln R\left(  \frac{1}{6}\ln^3\delta_2-\frac{1}{6}\pi^2\ln \delta_2 -3\zeta(3)     \right)   -\frac{1}{12}\pi^2\ln^2\delta_2\nn
\\& -\frac{\pi^4}{18}   \Bigg] \,.
\end{align}
The other integrals can be calculated in a similar method, and we obtain the amplitude in this region:
\begin{align}
    \mathcal{A}^{(1)a}_{(c-c),l_T\le m_H}=&\left(     \frac{\mu^2}{m_b^2}   \right)^{2\epsilon} \Bigg[   \frac{1}{\epsilon}\Bigg(    \ln R \left(2 \ln ^2\delta_2+4 \ln \delta_2+\frac{4 \pi ^2}{3}\right)    \Bigg) \nn
    \\&
     +\ln ^2R \left(2 \ln ^2\delta_2+4 \ln \delta_2+\frac{4 \pi ^2}{3}\right)+\ln R \left(\frac{2}{3} \ln ^3\delta_2+2 \ln ^2\delta_2\right.\nn
     \\&\left.
     -\frac{2}{3} \pi ^2 \ln \delta_2+8 \ln \delta_2-12 \zeta (3)+\frac{2 \pi ^2}{3}\right)-\frac{1}{3} \pi ^2 \ln ^2\delta_2-\frac{2}{3} \pi ^2 \ln \delta_2\nn
     \\&
     -\frac{2 \pi ^4}{9}      \Bigg]\,.
\end{align}
The result of the $l$-collinear  sector is given by $\mathcal{A}^{(1)a}_{(c),l_T\le m_H}=\mathcal{A}^{(1)a}_{(c-c),l_T\le m_H}+\mathcal{A}^{(1)a}_{(c-h),l_T\le m_H}$.
The result of the anti-collinear sector can be obtained via replacing $\delta_2$ by $\delta_1$.

In the $l$-soft sector, we also perform the integration of the loop momentum $q$ in different regions firstly:
\begin{align}
q\text{-hard:} \quad &  \left(- \frac{\mu^2}{m_H^2} \right)^{\epsilon}\Bigg(  -\frac{1}{\epsilon^2}+\frac{\pi^2}{12}-1  \Bigg)\,, \nn\\
q\text{-hard-collinear:} \quad & \left( \frac{\mu^2}{m_H(n\cdot l)} \right)^{\epsilon}\Bigg(  \frac{1}{\epsilon^2}+\frac{1}{\epsilon} -\frac{\pi^2}{12} +2 \Bigg)\,, \nn\\
q\text{-hard-anti-collinear:} \quad &  \left( -\frac{\mu^2}{m_H(\bar n\cdot l)} \right)^{\epsilon}\Bigg(  \frac{1}{\epsilon^2}+\frac{1}{\epsilon} -\frac{\pi^2}{12} +2 \Bigg)\,, \nn\\
q\text{-soft:} \quad &  \left( \frac{\mu^2}{(n\cdot l)(\bar n\cdot l)} \right)^{\epsilon}\Bigg(  -\frac{1}{\epsilon^2}-\frac{\pi^2}{4} \Bigg)\,.
\end{align}
The hard-anti-collinear region gives the same result as the hard-collinear region with $m_H(n\cdot l)$ replaced by $-m_H(\bar{n}\cdot l)$.
We see that there are double poles in each individual region, though they cancel out in their sum.
It is interesting to compare with the results in the collinear sector where only single poles appear; see eq. (\ref{eq:qhard-res}) and eq. (\ref{eq:qcoll-res}) above.
However, this does not mean that the soft sector is more complicated than the collinear sector.
On the contrary, it is harder to calculate the fixed-order result of the collinear sector and to understand its all-order logarithmic structure.
In the above equations, the scale in each region is reflected by the non-analytic terms, e.g., $(-m_H^2)^{-\epsilon}$.
In contrast, the non-analytic terms in eq. (\ref{eq:qhard-res}) and eq. (\ref{eq:qcoll-res}) are not so simple.

The remaining $l$ integration can be performed in a standard method, though we have implemented {\tt{HyperInt}} \cite{Panzer:2014caa} and used the relations among MPLs in \cite{Frellesvig:2016ske} to deal with some specific integrals.
When the dust settles, the result of the soft sector is given by
\begin{align}
\mathcal{A}^{(1)a}_{(s),l_T\le m_H}=&\mathcal{A}^{(1)a}_{(s-s),l_T\le m_H}+\mathcal{A}^{(1)a}_{(s-hc),l_T\le m_H}+\mathcal{A}^{(1)a}_{(s-h\bar c),l_T\le m_H}+\mathcal{A}^{(1)a}_{(s-h),l_T\le m_H}\nn
\\= &  \left(     \frac{\mu^2}{m_H^2}   \right)^{2\epsilon}     \left[   \frac{1}{\epsilon}    \left(    -2\ln^{2}R+\ln R\left(8\ln\delta_2-4i\pi\right)   \right) + \frac{\ln^{4}R}{12}+\ln^{3}R\left(2+\frac{i\pi}{3}\right)\right.\nn
\\&\left.-\ln^{2}R\left(\ln^{2}\delta_2  +6\ln\delta_2+\frac{\pi^{2}}{2}-2i\pi + 3\right)+\ln R\left(\frac{4}{3}\ln^{3}\delta_2-2i\pi\ln^{2}\delta_2\right.\right.\nn
\\&\left.\left.  +4i\pi\ln\delta_2+12\ln\delta_2+\frac{8\pi^{2}}{3}      -\frac{i\pi^{3}}{3}-6i\pi\right)-\frac{4}{3} \pi ^2 \ln\delta_2+4\zeta(3)+\frac{2i\pi^{3}}{3}  \right] \nn
\\&  + (\delta_2 \to \delta_1)\,,
\end{align}

Summarizing the above results in the collinear and soft sectors, we obtain
\begin{align}
\mathcal{A}^{(1)a}_{l_T\le m_H}=&\mathcal{A}^{(1)a}_{( c),l_T\le m_H}+\mathcal{A}^{(1)a}_{(\bar c),l_T\le m_H}-\mathcal{A}^{(1)a}_{(s),l_T\le m_H}\nn
\\=&  \left(     \frac{\mu^2}{m_H^2}   \right)^{2\epsilon}     \left[   \frac{1}{\epsilon} \left(      4\ln^2 R + 8 i\pi\ln R     \right) -\frac{1}{6}\ln^{4}R-\ln^{3}R\left(4+\frac{2i\pi}{3}\right)    \right.\nn
\\&\left. -\ln^{2}R\left( \frac{\pi^{2}}{3}+ 4i\pi-6\right)-\ln R\left(32\zeta(3)+2i\pi^{3}+\frac{16\pi^{2}}{3}-12i\pi\right) - 8\zeta(3)
\right.\nn
\\&\left. - \frac{4i\pi^{3}}{3}   \right]\,,
\end{align}
where all the $\delta_i$ dependencies are canceled, as expected.
The above result reproduces all the logarithmic terms $\ln^i R$ with $i=1,2,3,4$ in eq. (\ref{eq:expand_NLOa}) which is calculated in the full theory.
We see that the $\ln^4 R $ and $\ln^3 R$ terms are solely from the soft sectors.
This is a feature that persists at all orders, i.e., the $\alpha_s^n \ln^{2n+2} R$ and $\alpha_s^n \ln^{2n+1} R$ logarithmic terms can be derived from the soft sector by taking $\delta_i=1$ after obtaining the analytic result of this sector.

\subsubsection{Diagram \ref{fig:two_loop_dia}(b)}
\begin{figure}[H]
  
 \centering
 \vspace{-0.3cm}
 \subfloat[]
{\includegraphics[width=0.25\textwidth]{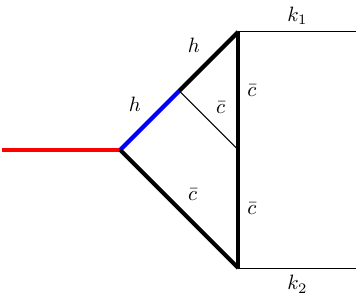}}
\subfloat[]
{\includegraphics[width=0.25\textwidth]{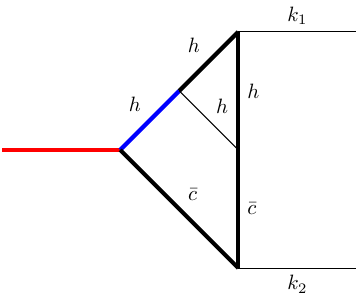}}
 \subfloat[]
{\includegraphics[width=0.25\textwidth]{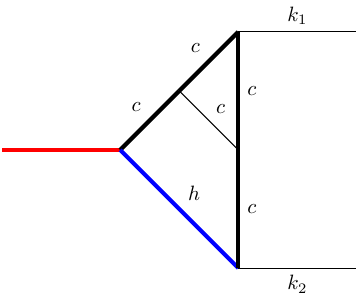}}

\subfloat[]
{\includegraphics[width=0.25\textwidth]{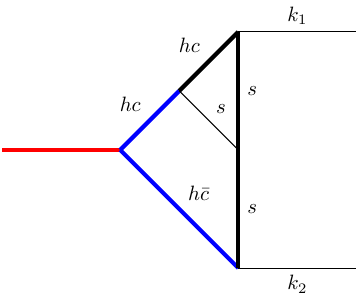}}
 \subfloat[]
{\includegraphics[width=0.25\textwidth]{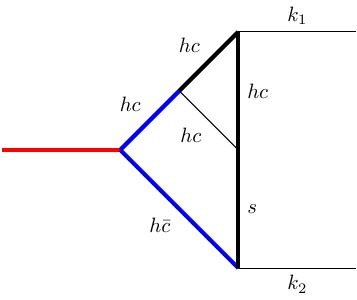}}
\caption{Regions contributing to LP logarithms for the diagram \ref{fig:two_loop_dia}(b) when we use the $\Delta$ regulator. The propagator with a $\Delta$ is in blue. }
\label{fig:dia_b_region_delta}
\end{figure}
The regions of the diagram \ref{fig:two_loop_dia}(b)  that can contribute to LP logarithms are shown in figure \ref{fig:dia_b_region_delta}.
The integrations of the loop momentum $q$ in the $(\bar c-\bar c)$ and $(s-s)$  regions are the same, given by
\begin{align}
   \int \frac{d^Dq}{(2\pi)^D}\frac{1}{(q^2-m_b^2)(l-q)^2} &=\frac{i(4\pi)^{\epsilon-2} \Gamma(\epsilon)m_b^{-2\epsilon}}{1-\epsilon} {}_2F_1\left(1,\epsilon;2-\epsilon;\frac{l^2}{m_b^2}+i0\right)
   \label{eq:B0fun}\\
   &=\frac{i(4\pi)^{\epsilon-2}\Gamma(\epsilon) (m_b^2-l^2)^{-\epsilon}}{1-\epsilon} {}_2F_1\left(1-\epsilon,\epsilon;2-\epsilon;\frac{l^2}{l^2-m_b^2+i0}\right)\nn
\end{align}
where $_2F_1$ is the hypergeometric function. 
Note that the hard (collinear) propagator has been canceled by the numerator.
Although the above two expressions are equivalent, it is important to use the second form to see clearly that the integration in the region of $l_T^2>m_H^2$ does not introduce any logarithms of $m_H^2/m_b^2$ at LP\footnote{At next-to-leading power, there may be $m_b^2\ln (m_H^2/m_b^2)$ terms in the result of the integral.}. 
After integration over $l$, we obtain
\begin{align}
\mathcal{A}^{(1)b}_{(\bar c- \bar c),l_T\le m_H}=& \left(     \frac{\mu^2}{m_H^2}   \right)^{2\epsilon}  \left[         \frac{4\ln R\ln\delta_1}{\epsilon}       -4\ln^{2}R\ln\delta_1+     \ln R\left(8\ln\delta_1+\frac{2\pi^{2}}{3}\right)+4\zeta(3)       \right]\,, \nn \\
\mathcal{A}^{(1)b}_{(s-s),l_T\le m_H}=& \left(     \frac{\mu^2}{m_H^2}   \right)^{2\epsilon}  \left[         \frac{1}{\epsilon}   \left(    -2\ln^{2}R+\ln R\left(4\ln\delta_1+4\ln\delta_2-4i\pi\right)   \right)       \right. \nn 
\\& \left. +\frac{8\ln^{3}R}{3}+\ln^{2}R\left(-4\ln\delta_1-4\ln\delta_2-4 + 4i\pi\right)\right. \nn 
\\& \left. +\ln R\left(8\ln\delta_1+8\ln\delta_2-8i\pi+\frac{2\pi^{2}}{3}\right)    \right]\,.
\end{align}

Similarly, the $(\bar c-h)$ and  $(s-h c)$ regions give the same result of the $q$ loop integration:
\begin{align}
   &\int \frac{d^Dq}{(2\pi)^D}\frac{-2\epsilon q^2+2(n\cdot q)(\bar n\cdot q)-2m_H( n\cdot q)}{q^2[q^2-( n\cdot l)(\bar n\cdot q)][q^2-m_H( n\cdot q)]} \nn
    \\=&\frac{i (1+ 2 \epsilon ) \Gamma (1-\epsilon ) \Gamma (-\epsilon ) \Gamma (\epsilon )(4\pi)^{\epsilon} \big(m_H( n\cdot l)-i0 \big)^{-\epsilon}}{32 \pi ^2 (-1+2 \epsilon ) \Gamma (-2 \epsilon )}\,.
\end{align}
The integration of $l$ can be carried out following the same method as in the one-loop integral, leading to the results
\begin{align}
\mathcal{A}^{(1)b}_{(\bar c- h),l_T\le m_H}=& \left(     \frac{\mu^2}{m_H^2}   \right)^{2\epsilon}  \left[         \frac{-2\ln R\ln\delta_1}{\epsilon}  + \ln^{2}R\ln\delta_1   \right. \nn 
\\& \left.   + \ln R\left(\ln^{2}\delta_1-2i\pi\ln\delta_1-8\ln\delta_1+\frac{\pi^{2}}{3}\right)+\frac{1}{3}\pi^{2}\ln\delta_1      \right]\,,\\
\mathcal{A}^{(1)b}_{(s-h c),l_T\le m_H}=& \left(     \frac{\mu^2}{m_H^2}   \right)^{2\epsilon}  \left[         \frac{1}{\epsilon}   \left(   \ln^{2}R-\ln R\left(2\ln\delta_1 + 2\ln\delta_2-2i\pi\right)  \right)       \right. \nn 
\\& \left.       -\ln^{3}R+\ln^{2}R\left(2\ln\delta_2+\ln\delta_1+4 - i\pi\right)+\ln R\left(-\ln^{2}\delta_2      \right.\right. \nn 
\\& \left. \left.     + \ln^{2}\delta_1-8\ln\delta_2-2i\pi\ln\delta_1-8\ln\delta_1-\frac{4\pi^{2}}{3}+8i\pi\right)    \right. \nn 
\\& \left.   +\frac{1}{3}\pi^{2}\ln\delta_2  +\frac{1}{3}\pi^{2}\ln\delta_1-2\zeta(3)-\frac{i\pi^{3}}{3}      \right]\,.
\end{align}

\begin{figure}[H]
 \centering
 \vspace{-0.3cm}
 \subfloat[]
{\includegraphics[width=0.335\textwidth]{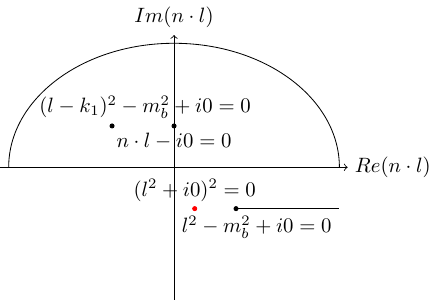}}
\subfloat[]
{\includegraphics[width=0.335\textwidth]{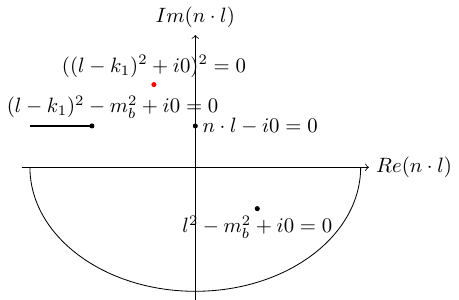}}
\subfloat[]
{\includegraphics[width=0.335\textwidth]{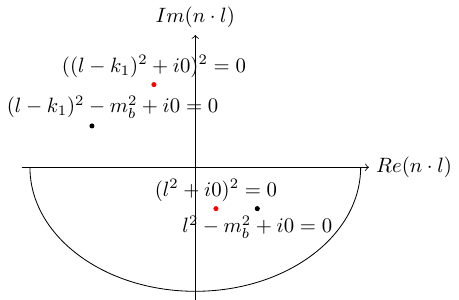}}
\caption{ The pole structures and branch cuts in the $(c-c)$ region. The black and red dots represent poles of order one and two, respectively. The thick line denotes a branch cut. Our integration contours are also shown by the arcs.}
\label{fig:branch_cut}
\end{figure}

The integration of the loop momentum $q$ in the $(c-c)$ region generates a lengthy result that is not appropriate to be shown explicitly.
It brings new pole structures like
\begin{align}
    \frac{1}{l^2+i0}, \quad \frac{1}{(l-k_1)^2+i0}, \quad \frac{1}{n\cdot l - i0}
\end{align}
and new branch cuts via the functions 
\begin{align}
    \ln (-l^2 + m^2 - i0),  
    \ln (-(l-k_1)^2 +m^2 - i0),
    \text{Li}_2\left( \frac{l^2}{m^2} +i0\right), 
    \text{Li}_2\left( \frac{(l-k_1)^2}{m^2} +i0 \right)
\end{align}
to the remaining integral of $l$.
Application of the residue theorem in the integration of $n\cdot l$, in which we choose the contours shown in figure \ref{fig:branch_cut},  implies $0<\bar n\cdot l<m_H$.
Then we integrate over $\bar{n}\cdot l$ and $l_T^2$ using {\tt HyperInt}.
In our calculation, we have expanded the result in $\Delta_2$ before performing the integration of $l_T^2$.
It is also important to collect all the terms from different integration contours in figure \ref{fig:branch_cut}
so that the unphysical poles of $1/l_T^2$ cancel in their sum.
Finally, we obtain
\begin{align}
\mathcal{A}^{(1)b}_{( c-  c),l_T\le m_H}=& \left(     \frac{\mu^2}{m_H^2}   \right)^{2\epsilon}  \left[         \frac{2\ln R\ln\delta_2}{\epsilon}  -2\ln^{2}R\ln\delta_2-\ln R\left(\ln^{2}\delta_2+\frac{5\pi^{2}}{3}\right)  \right.\nn
\\& \left.   +\frac{1}{3}\pi^{2}\ln\delta_2-4\zeta(3)-\frac{\pi^{2}}{3}    \right]\,.
\end{align}

Summarising the above results, we obtain
\beq
\begin{aligned}
\mathcal{A}^{(1)b}_{l_T\le m_H} =&\mathcal{A}^{(1)b}_{(\bar c- \bar c),l_T\le m_H}+\mathcal{A}^{(1)b}_{(\bar c- h),l_T\le m_H}+\mathcal{A}^{(1)b}_{( c-  c),l_T\le m_H}-\mathcal{A}^{(1)b}_{(s-s),l_T\le m_H}-\mathcal{A}^{(1)b}_{(s-h c),l_T\le m_H}
\\=& \left(     \frac{\mu^2}{m_H^2}   \right)^{2\epsilon}  \left[         \frac{1}{\epsilon}\left(   \ln^{2}R + 2i\pi\ln R     \right)   -\frac{5}{3}\ln^{3}R - 3i\pi\ln^{2}R + 2\zeta(3)+\frac{i\pi^{3}}{3}-\frac{\pi^{2}}{3}   \right]\,,
\end{aligned}
\eeq
where all the $\delta_i$ dependencies are canceled, as expected.
The above result reproduces all the logarithmic terms $\ln^i R$ with $i=1,2,3$ in eq. (\ref{eq:expand_NLOb}) which is calculated in the full QCD.
Again, the $\ln^3 R$ term comes only from the $l$-soft sector.

\subsubsection{Diagrams \ref{fig:two_loop_dia}(d)(e)}

\begin{figure}[t]
 \centering
 \vspace{-0.3cm}
\subfloat[]
{\includegraphics[width=0.25\textwidth]{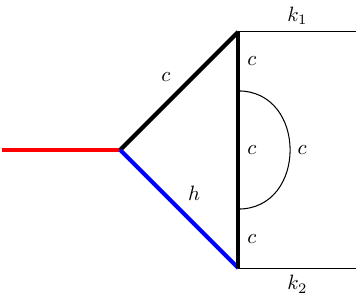}}
 \subfloat[]
{\includegraphics[width=0.25\textwidth]{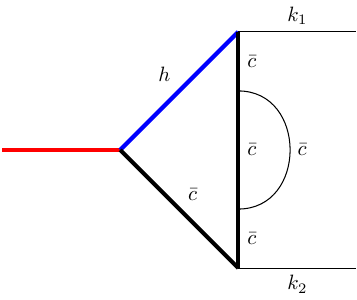}}
 \subfloat[]
{\includegraphics[width=0.25\textwidth]{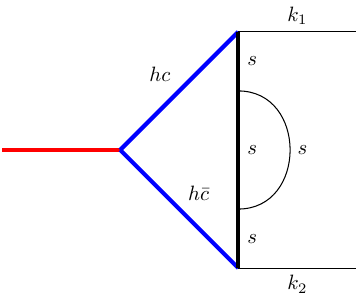}}

\subfloat[]
{\includegraphics[width=0.25\textwidth]{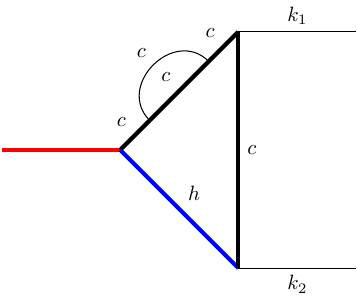}}
 \subfloat[]
{\includegraphics[width=0.25\textwidth]{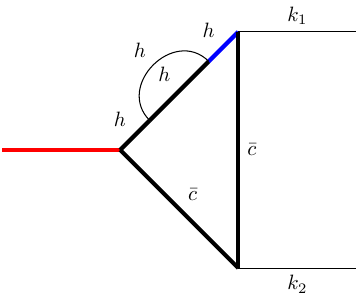}}
 \subfloat[]
{\includegraphics[width=0.25\textwidth]{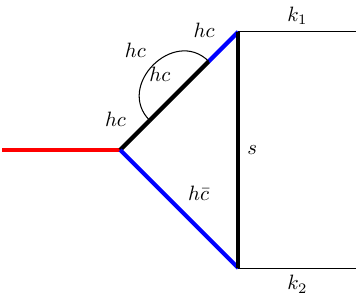}}
\caption{ Regions contributing to LP logarithms for the diagrams \ref{fig:two_loop_dia}(d) (upper plots) and \ref{fig:two_loop_dia}(e) (lower plots) when we use the $\Delta$ regulator. The propagator with a $\Delta$ is in blue. }
\label{fig:dia_d_region_delta}
\end{figure}

For the diagram \ref{fig:two_loop_dia}(d), the regions that can contribute to LP are shown in figure \ref{fig:dia_d_region_delta}.
As in eq. (\ref{eq:B0fun}), the integration of the loop momentum $q$ can be performed in arbitrary spacetime dimension, 
and the results are written in terms of hypergeometric functions,
which would introduce branch cuts at $l^2-m^2=0$. 
Applying the method described in detail above to the remaining $l$ integration, we obtain the following results:
\begin{align}
\mathcal{A}^{(1)d}_{(c-c),l_T\le m_H}=& \left(     \frac{\mu^2}{m_H^2}   \right)^{2\epsilon}  \left[         \frac{1}{\epsilon}\left(   \ln R\left(4\ln\delta_2+6\right)+12\ln\delta_2+12   \right)  \right.\nn
\\&\left.      -\ln^{2}R\left(4\ln\delta_2+6\right)+\ln R\left(-16\ln\delta_2+\frac{2\pi^{2}}{3}-20\right)-\frac{4}{3}\pi^{2}\ln\delta_2 \right.\nn
\\&\left.   +16\ln\delta_2-4\zeta(3)-\frac{2\pi^{2}}{3}+20     \right] \,, 
\label{eq:Adcc}\\
\mathcal{A}^{(1)d}_{(s-s),l_T\le m_H}=& \left(     \frac{\mu^2}{m_H^2}   \right)^{2\epsilon}  \Bigg[         \frac{1}{\epsilon}\bigg(   -\ln^{2}R+\ln R\left(4\ln\delta_2-6 - 2i\pi  \bigg)     +12 \ln \delta_2  -6 i \pi   \right)    \nn
\\& +\frac{4}{3}\ln^{3}R+  \ln^{2}R\left(-     4\ln\delta_2+10 + 2i\pi\right)+\ln R\left(-16\ln\delta_2-8 + 8i\pi + \pi^{2}\right)   \nn
\\&   -\frac{4}{3}\pi^{2}\ln\delta_2+16\ln\delta_2+4\zeta(3)+\frac{2i\pi^{3}}{3}    +\pi^{2}-8i\pi      \Bigg] +(\delta_2 \to \delta_1) \,.
\end{align}
The $(\bar c-\bar c)$ region gives the same result as in  eq. (\ref{eq:Adcc}) with $\delta_2$ replaced by $\delta_1$.
Summarising the above results, we obtain
\begin{align}
\mathcal{A}^{(1)d}_{l_T\le m_H} =&\mathcal{A}^{(1)d}_{(c-c),l_T\le m_H}+\mathcal{A}^{(1)d}_{(\bar c- \bar c),l_T\le m_H}-\mathcal{A}^{(1)d}_{( s- s),l_T\le m_H}
\nn \\=& \left(     \frac{\mu^2}{m_H^2}   \right)^{2\epsilon}  \bigg[         \frac{1}{\epsilon}\left(     2\ln^{2}R+  \ln R(24 + 4i\pi) + 12i\pi + 24      \right)       -\frac{8}{3}\ln^{3}R  \quad \quad \quad \quad \quad \quad  
\nn \\& -\ln^{2}R(32 + 4i\pi)       -  \ln R\left(\frac{2\pi^{2}}{3} + 16i\pi +24 \right) - 16\zeta(3)-\frac{4i\pi^{3}}{3}-\frac{10\pi^{2}}{3} 
\nn \\&+16i\pi + 40     \bigg] \,,
\end{align}
which agrees with eq. (\ref{eq:expand_NLOd}) for the  logarithmic terms at LP.
Note that the $\ln^3 R$ term comes only from the $l$-soft sector.

The calculation for the diagram \ref{fig:two_loop_dia}(e) is similar, and we get
\begin{align}
\mathcal{A}^{(1)e}_{l_T\le m_H} =& \left(     \frac{\mu^2}{m_H^2}   \right)^{2\epsilon}  \left[         \frac{ 1 }{\epsilon} \left(  -\ln^{2}R - \ln R(6 + 2i\pi) - 12    \right) +         \ln^{3}R+\ln^{2}R(5 + i\pi) \right.\nn
\\&\left. +\ln R\left(\frac{4\pi^{2}}{3}  - 2i\pi+ 20 \right) + 22\zeta(3)+\frac{i\pi^{3}}{3}+\pi^{2}-20  \right] 
\end{align}
which reproduces all the logarithmic terms in eq. (\ref{eq:expand_NLOe}) at LP.

\section{Implication for the factorization and resummation}
\label{sec:discussion}

\begin{figure}[H]
 \centering
 \vspace{-0.3cm}
 \subfloat[]
{\includegraphics[width=0.3\textwidth]{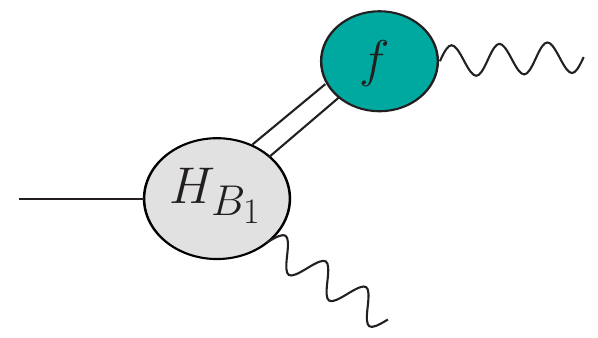}}
\subfloat[]
{\includegraphics[width=0.3\textwidth]{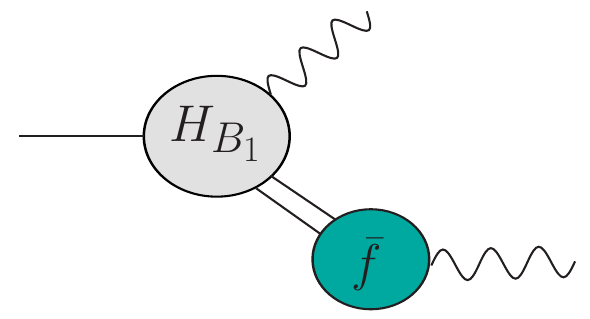}}
\subfloat[]
{\includegraphics[width=0.3\textwidth]{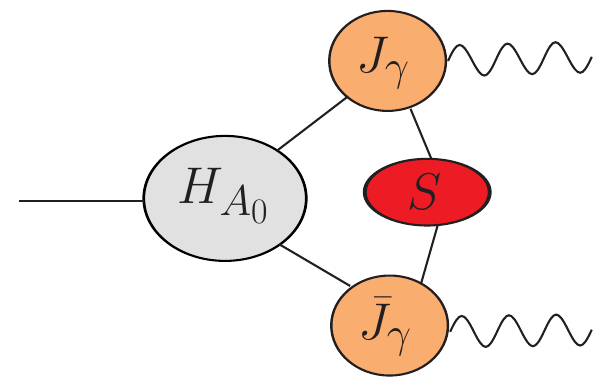}}
\caption{Graphical illustration of the NLP factorization.   }
\label{fig:factorization}
\end{figure}

Based on the above region analysis, we get a clear picture of the factorization formula for the decay amplitude, which can be power expanded as 
\begin{align}
   \mathcal{M}  & =  \mathcal{M}^{\rm NLP} + \mathcal{M}^{\rm NNLP} + \mathcal{O}(\lambda^6).
\end{align}
Here the subscript NLP and NNLP denote that the contributions are of order $\lambda^2$ and $\lambda^4$, respectively.
Making use of the analytical regulator, the NLP amplitude can be expressed as 
\begin{align}
   \mathcal{M}^{\rm NLP}  & = H_{A0}^{\gamma\gamma}(\mu)
   +\int_0^1 dx H_{B1}(\mu,\nu,x)  f_{b\bar{b}\to \gamma}(\mu,x) 
   +\int_0^1 dy H_{B1}(\mu,y) \bar{f}_{b\bar{b}\to \gamma}(\mu,\nu,y),
   \label{eq:ampNLP1}
\end{align}
where $H_{A0}^{\gamma\gamma}$ and $H_{B1}$ denote the contributions from the hard regions while $f_{b\bar{b}\to \gamma}$ and $\bar{f}_{b\bar{b}\to \gamma}$ represent the collinear contributions.
The first term corresponds to the diagrams in which all the propagators are of the hard scale; see, e.g., eq. (\ref{eq:A1ahh}).
The last two terms indicate the diagrams containing both a hard  and a collinear subgraphs.
The hard and collinear subgraphs are connected by two (anti-)collinear propagators. 
The structure is shown in the first two graphs in figure \ref{fig:factorization}.
The momentum fraction of the large light-cone component carried by one of the (anti-)collinear propagator is denoted by $x$ ($y$).
For example, one can take $x\equiv \bar{n}\cdot l /m_H $ in eq. (\ref{eq:AcLO}).
The boundary of the variable $x$ is determined by the singularity structure of the integrand.
With the analytic regulator, the hard function $H_{B1}(\mu,\nu,x)$ contains a regulated endpoint singularity, $x^{-1-\eta}$ in the collinear sector (the second term in the above equation).
After integration, it transforms to the divergences of $1/\eta$.
In the anti-collinear sector, the analytic regulator is imposed on the anti-collinear function and appears in the form of $y^{\eta}$, which regulates the endpoint singularity of $1/y$ in $H_{B1}(\mu,y)$.
The $1/\eta$ poles must cancel between the results of the collinear and anti-collinear sectors, as we have shown at two-loop level in section \ref{sec:expansion}.
This cancellation occurs actually even for each diagram.
This property may be an indispensable principle in developing a scheme to perform resummation of the logarithms to all orders.

\captionsetup[subfloat]{labelformat=empty}

From the calculation in the $\Delta$ regulator, 
we can write the NLP amplitude as
\begin{align}
   \mathcal{M}^{\rm NLP}  & = 
   \int_0^1 dx H_{B1}(\mu,\delta_2,x)  f^<_{b\bar{b}\to \gamma}(\mu,x) 
   +\int_0^1 dy H_{B1}(\mu,\delta_1,y) \bar{f}^<_{b\bar{b}\to \gamma}(\mu,y)\nn \\
   &-H_{A0}^{b\bar{b}} \int d\rho \int d\sigma J_{\gamma}(\rho,\delta_1)  \bar{J}_{\gamma}(\sigma,\delta_2)  S^<(\rho,\sigma)+\cdots.
      \label{eq:ampNLP2}
\end{align}
The first two terms are similar to the counterparts in eq. (\ref{eq:ampNLP1}).
However, the hard functions are now regulated by $\delta_i=\Delta_i/m_H^2$
and the collinear function is calculated with the constraint of $l_T<m_H$, which is indicated by the superscript $<$.
The term in the second line consists of the hard function $H_{A0}^{b\bar{b}}$, jet functions $J_{\gamma}(\rho,\delta_1)$ and $\bar{J}_{\gamma}(\sigma,\delta_2)$, and soft function $S^<(\rho,\sigma)$.
The structure of this term is illustrated in the last graph in figure \ref{fig:factorization}.
The endpoint divergences appear when integrating over $\rho$ and $\sigma$, which are the light-cone components of some loop momenta.
The ellipsis denotes the contribution of the pure hard region and the parts with $l_T>m_H$.
It would not affect the logarithms in the result, and thus is omitted in the above equation.

The factorization formulae in eqs. (\ref{eq:ampNLP1},   \ref{eq:ampNLP2}) are consistent with the region analysis carried out in section \ref{sec:expansion} and \ref{sec:Delta}.
In particular, no more momentum modes than those shown in eqs. (\ref{eq:modes1},\ref{eq:modes2}) are found.
Nonetheless, we have performed the calculation with Feynman parameters or loop-by-loop.
Therefore the explicit results of the various functions cannot be obtained directly from our calculations.
It is also promising to compute the functions from their definitions in the framework of an effective field theory.
We address that this factorization is different from that in refs. \cite{Liu:2019oav,Liu:2020wbn}, although they share the same structure.
Firstly, we do not rearrange the factorization formula to subtract the endpoint divergence,
and thus no additional operators in the endpoint region are introduced in the factorization and renormalization.
Secondly, since we have imposed the constraint on the transverse momentum, the leading order results are finite in each term of the factorization formula.
Consequently, no renormalization mixing between different operators is expected.
These features indicate that a compact resummation formula seems promising.
We leave these tasks to future work.

\begin{figure}[t]
 \centering
 \vspace{-0.3cm}
 \subfloat[(a)]
{\includegraphics[width=0.33\textwidth]{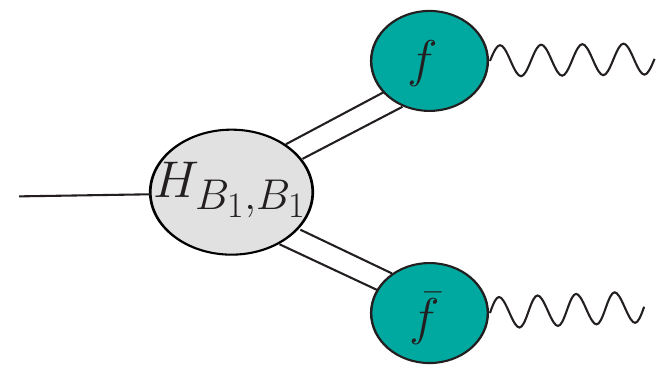}}
\subfloat[(b)]
{\includegraphics[width=0.33\textwidth]{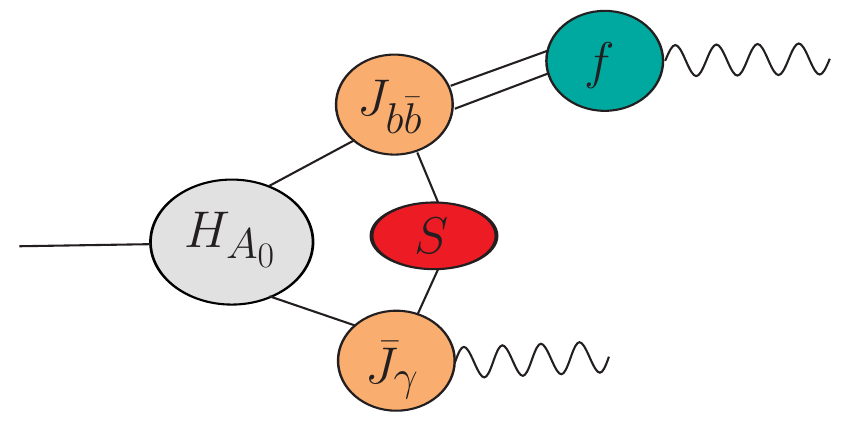}}
\subfloat[(c)]
{\includegraphics[width=0.33\textwidth]{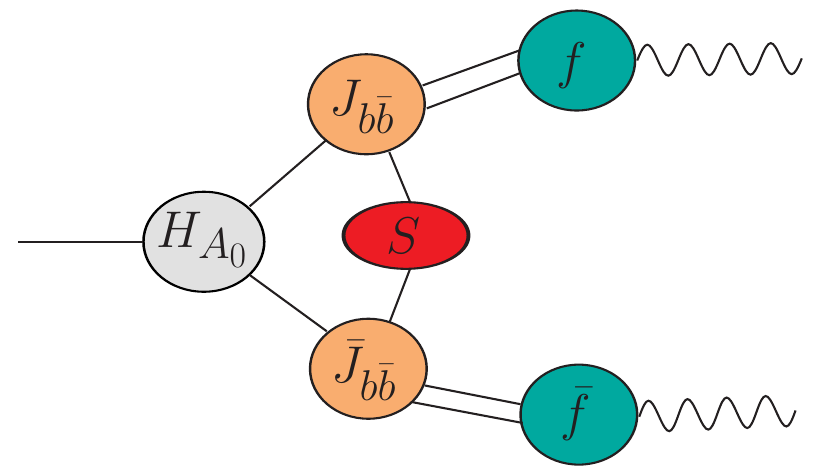}}
\caption{Graphical illustration of the NNLP factorization. }
\label{fig:factorization_NLP}
\end{figure}

Lastly, let us give a glimpse of the structure of the NNLP amplitude based on our region analysis.
Firstly, $\mathcal{M}^{\rm NNLP}$ receives contributions from the same regions as NLP.
One just needs to expand the integrand in the relevant region to higher power terms.
This kind of contribution can be considered as the insertion of higher power hard functions or soft-collinear interaction vertices in the diagrams in figure \ref{fig:factorization}.
Secondly, new topological contribution would appear.
The following factorization structure starts at NNLP\footnote{We omitted those structures with a massive tadpole subgraph, which is simple to calculate.}:
\begin{subequations}
 \begin{align}
   &  \int dx dy H_{B1,B1}(xy) f_{b\bar{b}\to \gamma}(x)  \bar{f}_{b\bar{b}\to \gamma}(y),
   \\
   & H_{A0}^{b\bar{b}} \int dx d\rho d\sigma   J_{b\bar b}(x\rho)  \bar{J}_{\gamma}(\sigma) S(\rho,\sigma)  f_{b\bar{b}\to \gamma}(x) ,\label{eq:facNNLP2}
   \\
   & H_{A0}^{b\bar{b}} \int dxdy d\rho d\sigma   J_{b\bar b}(x\rho)  \bar{J}_{b\bar b}(y\sigma) S(\rho,\sigma)  f_{b\bar{b}\to \gamma}(x) \bar{f}_{b\bar{b}\to \gamma}(y)    .\label{eq:facNNLP3}
\end{align}
\end{subequations}
The graphical illustration of the NNLP factorization is shown in figure \ref{fig:factorization_NLP}.
The hard function $H_{B1,B1}(xy)$ contains an endpoint singularity at $xy=0$.
Similarly, the jet function $J_{b\bar{b}}(x\rho)$ diverges at $x\rho=0$.
The cancellation of these two-dimensional endpoint singularities requires more intricate interplay among 
different regions than the one-dimensional endpoint singularity at NLP.
Note that the leading contribution of the factorization formula in eq. (\ref{eq:facNNLP2}) is of order $\lambda^3$ and will cancel with the other $\mathcal{O}(\lambda^3)$ expansion of the NLP amplitude.
Only the $\mathcal{O}(\lambda^4)$ result will survive and contribute to the NNLP amplitude.

\captionsetup[subfloat]{labelformat=empty}

It is also interesting to compare the NNLP factorization structure with that in electron-muon backward scattering; see (3.10) of ref. \cite{Bell:2022ott}.
Although there are many similarities, the differences are also clear.
First, our collinear function $f_{b\bar{b}\to \gamma}(x)$ transforms two collinear quarks to a collinear photon, and thus does not have the simple LO result $\delta(1-x)$.
This leads to the fact that the factorization diagrams in figure \ref{fig:factorization_NLP}(a,b) and (c)  appear from the two-loop and three-loop orders, respectively.
Second, we have a hard function in eq. (\ref{eq:facNNLP3}) and the soft function consists of only a single soft fermion line,
while no hard function exists and the soft function contains two soft fermion lines in ref. \cite{Bell:2022ott}.
We believe that the NNLP structure of the $H\to \gamma\gamma$ amplitude exhibits general features of the factorization formula beyond LP, deserving a thorough investigation in future.

\section{Conclusions}
\label{sec:conclusion}

We have presented a region analysis of the $H\to\gamma\gamma$ amplitude up to two-loop level.
The analysis is conducted using two kinds of regulators.
In the analytic regulator, the power of a propagator is increased from 1 to $1+\eta$.
All the propagators can  behave only in hard or (anti-)collinear modes. 
The endpoint singularities are encoded in $1/\eta$ poles,
which cancel between the collinear and anti-collinear sectors, though the results in the two sectors are not symmetric due to the presence of the analytic regulator.
An interesting feature is that the endpoint singularities may disappear in some two-loop diagrams.
This is caused by the specific loop structure.
It is also possible the endpoint singularities are regulated by the dimensional parameter at two-loop level.
In these diagrams, the $\eta$ regulators are not needed any more.

In the $\Delta$ regulator, the propagators that can potentially lead to endpoint singularities are modified by adding a $\Delta$ term.
Then the endpoint singularities emerge as $\ln^i(\Delta/m_H^2)$, which are clearly distinguished from the conventional infrared and ultraviolet divergences.
We have also split the bottom quark transverse momentum into two parts, i.e., $0<l_T<m_H$ and $l_T>m_H$.
The second part contributes only constants to the amplitude while all the logarithms come from the first part.
The important feature is that the LO results are finite and thus it is more convenient to study the relation between higher-order divergences and the associating logarithms.
Moreover, the symmetry between the collinear and anti-collinear sectors is preserved in this regulator, simplifying the calculation.
But the soft sector is not scaleless anymore.
Actually, the leading and next-to-leading logarithms can be derived only from the soft sector.

Our region analysis may help to develop consistent factorization and resummation schemes in different regulators.
We also give a brief discussion on the NNLP amplitude based on the region analysis, 
which exhibits more complicated factorization structure such as the two-dimensional endpoint singularities.
It is intriguing to study the regularization and cancellation of such endpoint singularities.

\section*{Acknowledgments}
We are grateful to Yao Ma for helpful discussion. 
This work was supported in part by the National Natural Science Foundation of China under grant No. 12005117, No. 12321005, No. 12375076 and the Taishan Scholar Foundation of Shandong province (tsqn201909011).

\appendix

\section{Topologies of two-loop master integrals and the complete two-loop result}
\label{sec:app1}

\begin{figure}[ht]
    \centering
	\begin{minipage}{0.17\linewidth}
		\centering
		\includegraphics[width=1\linewidth]{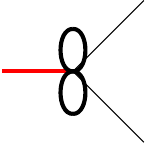}
		\caption*{$(1)$}
	\end{minipage}
    \centering
	\begin{minipage}{0.3\linewidth}
		\centering
		\includegraphics[width=1\linewidth]{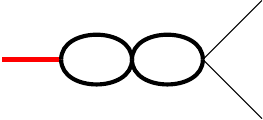}
		\caption*{$(2)$}
	\end{minipage}    
    \centering
	\begin{minipage}{0.23\linewidth}
		\centering
		\includegraphics[width=1\linewidth]{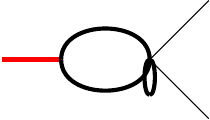}
		\caption*{$(3)$}
	\end{minipage}    
    \centering
	\begin{minipage}{0.2\linewidth}
		\centering
		\includegraphics[width=1\linewidth]{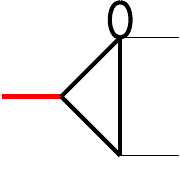}
		\caption*{$(4)$}
	\end{minipage}    
	\centering
	\begin{minipage}{0.23\linewidth}
		\centering
		\includegraphics[width=1\linewidth]{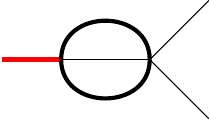}
		\caption*{$(5)$}
	\end{minipage}
    \centering
	\begin{minipage}{0.23\linewidth}
		\centering
		\includegraphics[width=1\linewidth]{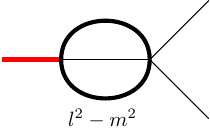}
		\caption*{$(6)$}
	\end{minipage}
    \centering
	\begin{minipage}{0.23\linewidth}
		\centering
		\includegraphics[width=1\linewidth]{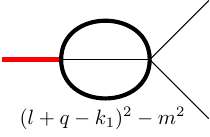}
		\caption*{$(7)$}
	\end{minipage}
    \centering
	\begin{minipage}{0.18\linewidth}
		\centering
		\includegraphics[width=1\linewidth]{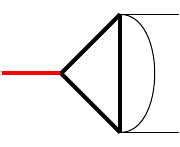}
		\caption*{$(8)$}
	\end{minipage}
    \centering
	\begin{minipage}{0.2\linewidth}
		\centering
		\includegraphics[width=1\linewidth]{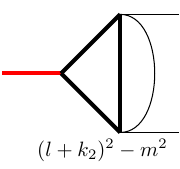}
		\caption*{$(9)$}
	\end{minipage}
    \centering
	\begin{minipage}{0.2\linewidth}
		\centering
		\includegraphics[width=1\linewidth]{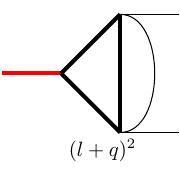}
		\caption*{$(10)$}
	\end{minipage}
    \centering
	\begin{minipage}{0.2\linewidth}
		\centering
		\includegraphics[width=1\linewidth]{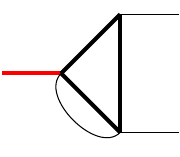}
		\caption*{$(11)$}
	\end{minipage}
    \centering
	\begin{minipage}{0.2\linewidth}
		\centering
		\includegraphics[width=1\linewidth]{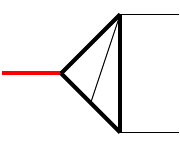}
		\caption*{$(12)$}
	\end{minipage}
    \centering
	\begin{minipage}{0.2\linewidth}
		\centering
		\includegraphics[width=1\linewidth]{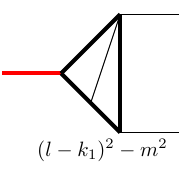}
		\caption*{$(13)$}
	\end{minipage}
    \centering
	\begin{minipage}{0.25\linewidth}
		\centering
		\includegraphics[width=1\linewidth]{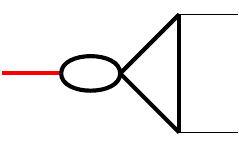}
		\caption*{$(14)$}
	\end{minipage}
\caption{The master integrals for the two-loop amplitude of  $H\to \gamma(k_1) \gamma(k_2)$ shown in figure \ref{fig:two_loop_dia}. The red lines represent Higgs boson, the thick lines represent massive propagators and the others are massless propagators. For the master integrals with $n_i<0$, we write the numerator at the bottom of the diagram.    }  
\label{fig:topo}
\end{figure}

The two-loop integral family can be represented as
\begin{align}
    I_{n_1,n_2,...,n_7}=\int \mathcal{D}^D l\mathcal{D}^D q \frac{D_7^{-n_7}}{D_1^{n_1}D_2^{n_2}D_3^{n_3}D_4^{n_4}D_5^{n_5}D_6^{n_6}}\,,
\end{align}
where $n_{1,...,6}\ge 0$, $n_7<0$, and the $D_i$ read
\begin{align}
    D_1&=l^2-m_b^2,&    D_2&=(l+k_2)^2-m_b^2,&     D_3&=(l-k_1)^2-m_b^2,& 
    \\D_4&=q^2,&   D_5&=(l+q+k_2)^2-m_b^2,     &D_6&=(l+q-k_1)^2-m_b^2,&  \ \  D_7&=(l+q)^2.&  \nn
\end{align}
This family contains fourteen master integrals, which are shown in figure \ref{fig:topo}.

The complete results for the two-loop diagrams are given by
\begin{align}
\mathcal{A}^{(1)a} =&\left( \frac{\mu^2}{m_b^2}\right)^{2\epsilon} \left[  \frac{1}{\epsilon}\left(\left(\frac{32}{z^2}-\frac{32}{z}+8\right) G(1,1,z)-16 \right)   +\left(\frac{40}{z^3}-\frac{60}{z^2}-\frac{44}{z}+32\right)G(1,z) \right.\nn
\\&+\left(-\frac{112}{z^3}+\frac{168}{z^2}-\frac{96}{z}+20\right)G(0,1,z)-\left(\frac{72}{z^4}-\frac{200}{z^3}+\frac{92}{z^2}+\frac{16}{z}+8\right)G(1,1,z)\nn
\\&+\left(-\frac{64}{z^2}+\frac{64}{z}-32\right)G(0,1,1,z)-\left(\frac{112}{z^4}-\frac{224}{z^3}+\frac{128}{z^2}-\frac{16}{z}\right)G(1,0,1,z)\nn
\\&+\left(\frac{56}{z^4}-\frac{16}{z^3}-\frac{16}{z^2}-\frac{24}{z}+24\right)G(1,1,1,z)+\left(-\frac{64}{z^2}+\frac{64}{z}-16\right)G(1,2,1,z)\nn
\\&+\left(-\frac{64}{z^2}+\frac{64}{z}\right)G(0,0,1,1,z)+\left(-\frac{32}{z^2}+\frac{32}{z}\right)G(0,1,0,1,z)\nn
\\&+\left(\frac{48}{z^2}-\frac{48}{z}\right)G(0,1,1,1,z)+\left(\frac{224}{z^3}-\frac{304}{z^2}+\frac{192}{z}-56\right)G(1,0,1,1,z)\nn
\\&+\left(-\frac{64}{z^3}+\frac{112}{z^2}-\frac{80}{z}+16\right)G(1,1,0,1,z)-\left(\frac{16}{z^3}-\frac{24}{z^2}+\frac{16}{z}-4\right)G(1,1,1,1,z)\nn
\\&\left.+\left(-\frac{256}{z^3}+\frac{384}{z^2}-\frac{256}{z}+64\right)G(1,1,2,1,z)\right.\nn
\\&\left.+\left(\frac{128}{z^3}-\frac{192}{z^2}+\frac{128}{z}-32\right)G(1,2,1,1,z)+\frac{20}{z^2} -\frac{20}{z} - 94 \right]\,,
\end{align}
\begin{align}
\mathcal{A}^{(1)b} =&\left( \frac{\mu^2}{m_b^2}\right)^{2\epsilon} \left[ \frac{1}{\epsilon}\left(\left(\frac{8}{z^2}-\frac{8}{z}+2\right) G(1,1,z)-4\right) +\left(-\frac{40}{z^3}+\frac{60}{z^2}-\frac{36}{z}+8\right)G(1,z) \right. \nn
\\&\left.+\left(\frac{112}{z^3}-\frac{168}{z^2}+\frac{80}{z}-12\right)G(0,1,z)+\left(\frac{72}{z^4}-\frac{200}{z^3}+\frac{204}{z^2}-\frac{88}{z}+12\right)G(1,1,z) \right.  \nn  \\&\left.+\left(-\frac{16}{z^2}+\frac{16}{z}+8\right)G(0,1,1,z)+\left(\frac{112}{z^4}-\frac{224}{z^3}+\frac{160}{z^2}-\frac{48}{z}+4\right)G(1,0,1,z) \right. \nn
\\&\left.+\left(-\frac{56}{z^4}+\frac{112}{z^3}-\frac{64}{z^2}+\frac{20}{z}-10\right)G(1,1,1,z)+\left(-\frac{16}{z^2}+\frac{16}{z}-4\right)G(1,2,1,z) \right. \nn
\\&\left.+\left(\frac{32}{z^2}-\frac{32}{z}\right)G(0,0,1,1,z)+\left(\frac{16}{z^2}-\frac{16}{z}\right)G(0,1,0,1,z) \right.\nn
\\&\left.+\left(-\frac{24}{z^2}+\frac{24}{z^2}\right)G(0,1,1,1,z)+\left(-\frac{16}{z^2}+\frac{16}{z^2}\right)G(1,0,1,1,z) \right.\nn
\\&\left.+\left(-\frac{8}{z^2}+\frac{8}{z}\right)G(1,1,0,1,z) -\frac{20}{z^2}+\frac{20}{z} - 22\right]\,,
\end{align}
\begin{align}
\mathcal{A}^{(1)d} =&\left( \frac{\mu^2}{m_b^2}\right)^{2\epsilon} \left[ -\frac{6}{\epsilon^2}   + \frac{1}{\epsilon}\left(  \left(\frac{192}{z^3}-\frac{288}{z^2}+\frac{120}{z}-12\right)G(1,z)      \right.\right.\nn
\\&+\left. \left.     \left(\frac{40}{z^2}-\frac{40}{z}+4\right)G(1,1,z)    +\frac{192}{z^2} -\frac{192}{z} +23     \right)       \right.\nn
\\&+\left.      \left(\frac{752}{z^3}-\frac{1128}{z^2}+\frac{420}{z}-22\right)G(1,z)      +\left(-\frac{112}{z^3}+\frac{168}{z^2}-\frac{64}{z}+4\right)G(0,1,z)    \right.\nn
\\&+\left.   \left(\frac{248}{z^3}-\frac{300}{z^2}+\frac{80}{z}-20\right)G(1,1,z)+\left(-\frac{384}{z^3}+\frac{576}{z^2}-\frac{240}{z}+24\right)G(2,1,z)    \right.\nn
\\&+\left.   \left(-\frac{32}{z^2}+\frac{32}{z}-8\right)G(0,1,1,z)+\left(-\frac{112}{z^4}+\frac{224}{z^3}-\frac{160}{z^2}+\frac{48}{z}\right)G(1,0,1,z)    \right.\nn
\\&+\left.   \left(\frac{56}{z^4}-\frac{112}{z^3}+\frac{136}{z^2}-\frac{80}{z}+8\right)G(1,1,1,z)+\left(-\frac{80}{z^2}+\frac{80}{z}-8\right)G(1,2,1,z)       \right.\nn
\\&\left.      +\frac{696}{z^2}-\frac{696}{z}-\pi^2+\frac{81}{2}          \right]\,,
\end{align}
\begin{align}
\mathcal{A}^{(1)e} =&\left( \frac{\mu^2}{m_b^2}\right)^{2\epsilon} \left[ \frac{3}{\epsilon^2} - \frac{1}{\epsilon}\left( \left(\frac{96}{z^3}-\frac{144}{z^2}+\frac{48}{z}\right)G(1,z)+\left(-\frac{4}{z^2}+\frac{4}{z}+2\right)G(1,1,z) +\frac{96}{z^2} \right.\right.\nn
\\&\left.\left. -\frac{96}{z} +\frac{23}{2} \right) + \left(-\frac{356}{z^3}+\frac{534}{z^2}-\frac{184}{z}+3\right)G(1,z) \right.\nn
\\&+\left. \left(-\frac{36}{z^4}-\frac{24}{z^3}+\frac{104}{z^2}-\frac{44}{z}-2\right)G(1,1,z)+\left(\frac{192}{z^3}-\frac{288}{z^2}+\frac{96}{z}\right)G(2,1,z)+ \right.\nn
\\&+\left.\left(\frac{80}{z^2}-\frac{80}{z}+32\right)G(0,1,1,z)+\left(-\frac{32}{z^2}+\frac{32}{z}-8\right)G(1,0,1,z)\right. \nn
\\&\left. + \left(\frac{12}{z^2}-\frac{12}{z}-6\right)G(1,1,1,z)+\left(-\frac{72}{z^2}+\frac{72}{z}-12\right)G(1,2,1,z) \right.\nn
\\&\left. -\frac{338}{z^2}+\frac{338}{z} - \frac{93}{4}+\frac{\pi^2}{2}\right] \,,
\end{align}
where 
\beq
\begin{aligned}
z=\frac{2}{\sqrt{1-4R}+1}\,.
\end{aligned}
\eeq

The renormalized two-loop correction reads
\begin{align}
\mathcal{A}^{(1)}=&\frac{1}{(\omega-1)^3}\left. \bigg(  1152\text {Li} _ 4 (-\omega) + 864\text {Li} _ 4 (\omega)  -  512\text {Li} _ 3 (-\omega)\ln (\omega) -  512\text {Li} _ 3 (\omega)\ln (\omega) \right.\nn
\\&\left.  + 64\text {Li} _ 2 (-\omega)\ln^2 (\omega) +  112\text {Li} _ 2 (\omega)\ln^2 (\omega)  + \frac {2\ln^4 (\omega)} {3} - 16\ln^3 (\omega) + \frac {16} {3}\pi^2\ln^2 (\omega)  \right.\nn
\\&\left. + 128\zeta (3)\ln (\omega) + \frac {8\pi^4} {5}    \right)   
+              \frac{1}{(\omega-1)^2}\left.\bigg(1296 \text{Li}_4(\omega )+ 1728 \text{Li}_4(-\omega ) \right.\nn
\\&\left.-768 \text{Li}_3(-\omega ) \ln (\omega )  -768 \text{Li}_3(\omega ) \ln (\omega ) +\ln ^4(\omega ) -256 \text{Li}_3(-\omega )  -96 \text{Li}_3(\omega ) \right.\nn
\\&\left. +96 \text{Li}_2(-\omega ) \ln ^2(\omega )+168 \text{Li}_2(\omega ) \ln ^2(\omega )+128 \text{Li}_2(-\omega ) \ln (\omega ) 
 +64 \text{Li}_2(\omega ) \ln (\omega )
 \right.\nn
\\&\left.-\frac{40 \ln ^3(\omega )}{3}+16 \ln (1-\omega ) \ln ^2(\omega )+  8 \pi ^2 \ln ^2(\omega )+24 \ln ^2(\omega )+192 \zeta (3) \ln (\omega )\right.\nn
\\&\left.-\frac{16}{3} \pi ^2 \ln (\omega )  +\frac{12 \pi ^4}{5}  -96 \zeta (3)   \right)     
+              \frac{1}{(\omega-1)}\left.\bigg(   1152 \text{Li}_4(-\omega )+864 \text{Li}_4(\omega )\right.\nn
\\&\left.+64 \text{Li}_2(-\omega ) \ln ^2(\omega )+     112 \text{Li}_2(\omega ) \ln ^2(\omega )-512 \text{Li}_3(-\omega ) \ln (\omega )-512 \text{Li}_3(\omega ) \ln (\omega )\right.\nn
\\&\left.+\frac{2 \ln ^4(\omega )}{3}-256\text{Li}_3(-\omega )-96\text{Li}_3(\omega ) -\frac{4}{3}\ln ^3(\omega ) +128\text{Li}_2(-\omega )\ln(\omega) \right.\nn
\\&\left.+64\text{Li}_2(\omega ) \ln(\omega)  +16\ln(1-\omega)\ln^2(\omega)+ \frac{16}{3} \pi ^2 \ln ^2(\omega )+24\ln^2(\omega)   \right.\nn
\\&\left.+128 \zeta (3) \ln (\omega )-\frac{16\pi^2}{3}\ln (\omega )+48\ln (\omega )+\frac{8 \pi ^4}{5} -96\zeta(3) \right) +288 \text{Li}_4(-\omega ) \nn
\\&+ 216 \text{Li}_4(\omega )+16 \text{Li}_2(-\omega ) \ln ^2(\omega ) +28 \text{Li}_2(\omega ) \ln ^2(\omega ) -128 \text{Li}_3(-\omega ) \ln (\omega )\nn
\\&-128 \text{Li}_3(\omega ) \ln (\omega )+\frac{\ln ^4(\omega )}{6}-64 \text{Li}_3(-\omega )-56 \text{Li}_3(\omega ) +32 \text{Li}_2(-\omega ) \ln (\omega )\nn
\\&+48 \text{Li}_2(\omega ) \ln (\omega )-2 \ln ^3(\omega ) +20 \ln (1-\omega ) \ln ^2(\omega )+\frac{4}{3} \pi ^2 \ln ^2(\omega )+32 \zeta (3) \ln (\omega )\nn
\\&-\frac{4}{3} \pi ^2 \ln (\omega )+24 \ln (\omega )+\frac{2 \pi ^4}{5}+8 \zeta (3)-40 \,.       
\end{align}
Our result is consistent with ref.~\cite{Harlander:2005rq}.

\bibliographystyle{JHEP}
\bibliography{ref}

\end{document}